\documentclass[twocolumn]{aastex63}

\usepackage{amsmath}

\accepted{March 2, 2021}
\submitjournal{ApJ}

\shorttitle{Braking MESA}
\shortauthors{Gossage et al.}
\graphicspath{{./}{figures/}}

\begin{document}

\title{MESA models with magnetic braking}

\correspondingauthor{Seth Gossage}
\email{sgossage@cfa.harvard.edu}

\author{Seth Gossage}
\affiliation{Center for Astrophysics | Harvard \& Smithsonian, 60 Garden Street, Cambridge, MA 02138, USA}

\author{Aaron Dotter}
\affiliation{Center for Astrophysics | Harvard \& Smithsonian, 60 Garden Street, Cambridge, MA 02138, USA}

\author{Cecilia Garraffo}
\affiliation{Institute for Applied Computational Science, Harvard University, 33 Oxford St., Cambridge, MA 02138, USA}

\author{Jeremy J. Drake}
\affiliation{Center for Astrophysics | Harvard \& Smithsonian, 60 Garden Street, Cambridge, MA 02138, USA}

\author{Stephanie Douglas}
\affiliation{Department of Physics, Lafayette College, 730 High St., Easton, PA 18042, USA}

\author{Charlie Conroy}
\affiliation{Center for Astrophysics | Harvard \& Smithsonian, 60 Garden Street, Cambridge, MA 02138, USA}




\begin{abstract}

Two magnetic braking models are implemented in \texttt{MESA} for use in the \texttt{MIST} stellar model grids. Stars less than about 1.3 solar masses are observed to spin down over time through interaction with their magnetized stellar winds (i.e., magnetic braking). This is the basis for gyrochronology, and fundamental to the evolution of lower mass stars. The detailed physics behind magnetic braking are uncertain, as are 1D stellar evolution models. Thus, we calibrate our models and compare to data from open clusters. Each braking model tested here is capable of reproducing aspects of the data, with important distinctions; neither fully accounts for the observations. The \cite{matt.etal:2015} prescription matches the slowly rotating stars observed in open clusters, but tends to overestimate the presence of rapidly rotating stars. The \cite{garraffo.etal:2018} prescription often produces too much angular momentum loss to accurately match the observed slow sequence for lower mass stars, but reproduces the bimodal nature of slow and rapidly rotating stars observed in open clusters fairly well. Our models additionally do not reproduce the observed solar lithium depletion, corroborating previous findings that effects other than rotation may be important. We find additional evidence that some level of mass dependency may be missing in these braking models to match the rotation periods observed in clusters older than 1 Gyr better.


\end{abstract}



\section{Introduction} \label{sec:intro}

Stellar rotation rate is one of the fundamental properties of stars, determining all other properties of a star throughout their evolution, alongside stellar mass and metallicity. Rotation introduces centrifugal forces that strongly affect the structure of a star, altering temperatures and luminosities (an effect called gravity darkening), as well as inducing mechanical instabilities that induce mixing of stellar material throughout the star \citep{vonzeipel:1924,endal.sofia:1976,heger.langer.woosley:2000,maeder.meynet:2000a,maeder:2009}. Due to the immense challenges of full, 3D stellar modeling, we rely on simpler 1D models. The effects of stellar rotation have been studied for over a century at least. Models exploring stellar rotation in 1D have been calculated over the last several decades, e.g., \cite{pinsonneault.etal:1989,meynet.maeder:2000,heger.langer:2000,palacios.etal:2003,charbonnel.talon:2005,denissenkov.pinsonneault:2007}. Numerous non-rotating grids (e.g., \citealt{baraffe.etal:1998,pietrinferni.etal:2004,demarque.etal:2004,dotter.etal:2008,bressan.etal:2012, choi.etal:2016}) with detailed physics and sometimes having their own specializations, e.g., as with the models of \cite{somers.pinsonneault:2015b,somers.cao.pinsonneault:2020} which models the structural effects magnetic fields and starspots, or e.g., binary evolution as in \cite{eldrige.etal:2017}. However, large 1D stellar model grids and isochrones that incorporate rotation effects over large ranges of mass, metallicity, and rotation rates, have only recently become available. This is not an exhaustive list, but see e.g., \citealt{ekstrom.etal:2012,georgy.etal:2014,gossage.etal:2018}. Until recently, grids that fully model 1D stellar rotation have been limited by a lack of models that describe the spin evolution of low mass stars (i.e., with about $\rm M<1.3 M_{\odot}$) in this context.

Implementing rotation in low mass 1D stellar models adds an additional layer of complexity. There are even fewer large grids of stellar models that incorporate the effects of rotation at low masses, but \cite{amard.etal:2019} has produced some of the first. At masses below about $1.3 \rm \rm M_{\odot}$, stars develop a convective envelope that increases in depth from the surface of the star until masses $\leq 0.3\ \rm \rm M_{\odot}$, below which stars become fully convective. The convective envelopes of these low mass stars host surface magnetic fields that interact with stellar winds, extracting angular momentum, and causing them to slow down over time. The magnetic coupling between star and stellar wind, leading to a slowing of the stellar rotation rate is commonly called magnetic braking. This phenomenon was inferred early on by \cite{kraft:1967} and \cite{skumanich:1972} in nearby stars, but was theoretically anticipated beforehand as in e.g., \cite{schatzman:1962,brandt:1966,weber.davis:1967}.

This characteristic of low mass stellar evolution may be exploited as an age determination technique known as gyrochronology (see e.g., \citealt{barnes:2003,mamajek.hillenbrand:2008}). Thus, the accuracy of this age determination method is partially reliant on an accurate modeling of rotation in low mass stars, including the effects of magnetic braking. Accurately modeling magnetic braking is also crucial to studies of stellar populations $\gtrsim 2$ Gyr old, when the main sequence turn off (MSTO) is dominated by low mass stars that experience magnetic braking \citep{georgy.etal:2019,gossage.etal:2019} and for studying their underlying rotation rate distributions. Furthermore, incorporating magnetic braking into 1D stellar models and applying them to data is a useful experiment to help constrain the detailed magnetohydrodynamic (MHD) simulations that inform them (e.g., \citealt{matt.etal:2012,matt.etal:2015,reville.etal:2015,garraffo.drake.cohen:2015,finley.matt:2018,garraffo.etal:2018,see.etal:2019}). Note also that the models we present in this work self-consistently evolve models with rotation effects, e.g., internal angular momentum transport and rotational mixing processes calculated (like \citealt{amard.etal:2019}). This is in contrast to other successful stellar models computed for gyrochronology that calculate rotation evolution from the structural properties of non-rotating stellar models (such as, \citealt{vansaders.pinsonneault:2013,gallet.bouvier:2013,gallet.bouvier:2015,lanzafame.spada:2015}), and thus do not self-consistently evolve the models with internal angular momentum transport, etc.

Furthermore, classical 1D stellar models can not reproduce the solar lithium abundance, and it has become clear that rotating stellar models, perhaps with additional physics (still under investigation) are required to explain it \citep{charbonnel.talon:2005,eggenberger.maeder.meynet:2005,somers.pinsonneault:2016}. The Sun has a greatly depleted surface abundance of lithium (measured at $\rm \log(^7Li/H)+12=1.05\pm0.1$ now, in comparison to its assumed primordial value $3.26\pm0.05$ measured from meteorites, e.g., \citealt{asplund.etal:2009}), and it is unclear why. Observations of solar twins \citep{carlos.etal:2019} and lithium abundances of stars in open clusters \citep{sestito.randich:2005} have shown that stars near solar mass generally show a depletion of surface lithium abundance; although, there is an unexplained dispersion in the abundance at a given age, and the Sun appears to be amongst the most heavily depleted compared to similar stars \citep{carlos.etal:2019}. It has been proposed that rotation-enhanced mixing could play a role in explaining this (e.g., \citealt{bouvier:2008,eggenberger.maeder.meynet:2010,somers.pinsonneault:2016}, and see \citealt{bouvier:2020} for an overview), and comparison of $v\sin i$ and lithium abundance data may support this connection, e.g., \cite{skumanich:1972,bouvier:2008,beck.etal:2017}. Lithium is not the only light element observed to experience depletion. Beryllium and boron are observed to be depleted as well in F- and B-type stars, simultaneously and in a similar fashion \citep{charbonnel.etal:1994,venn.etal:2002}. Thus, reproducing light element abundances like lithium can serve as a useful constraint on low mass, rotating stellar models, and help us understand what its cause may be.

Our models are based on the framework of the \texttt{MIST} stellar model grid \citep{dotter:2016,choi.etal:2016}, which was built to cover a large range of masses (0.1 to 300 $\rm M_{\odot}$), evolution from pre-main sequence (PMS) to late evolutionary phases (such as white dwarfs), and a wide range of metallicities. The current iteration of \texttt{MIST} contains only non-rotating models, and models at a single higher rotation rate; although \cite{gossage.etal:2018,gossage.etal:2019} have done work towards expanding the grid to include more rotation rates. All the while, \texttt{MIST} has never included a proper treatment of rotation for stars with surface convective envelopes. In this work, we implement two magnetic braking models: \cite{matt.etal:2015} and \cite{garraffo.etal:2018}. We examine their abilities to reproduce observed rotation periods of open cluster stars for a range of ages as implemented here. Our goal in doing this is to provide models available for studies, such as of the type listed above, but also to work towards converging our theories on the physics that drive magnetic braking. In so doing, hopefully we can make our 1D stellar models more accurate in this regime and uncover the physics behind rotation-driven phenomena. We have developed a preliminary grid of stellar models at solar metallicity to test implementations of magnetic braking in our models.

In Sec. \ref{sec:mesa}, we provide the background of our models and some of the fundamental framework behind stellar rotation in \texttt{MESA}. We describe our implementation of the \cite{matt.etal:2015} and \cite{garraffo.etal:2018} braking models in Sec. \ref{sec:blaws}. Our results are presented in Sec. \ref{sec:results}, displaying the effects of each braking model in regards to reproducing observed open cluster rotation periods, effects on the Hertzsprung-Russell diagram, and surface lithium depletion. We discuss our results and discrepancies found in Sec. \ref{sec:discussion}. Finally, we conclude our work in Sec. \ref{sec:conclusions}.  


\section{Rotation in MESA} \label{sec:mesa}

Modeling the effects of stellar rotation on the evolution of a star is difficult. It is at least a 2D effect, but with current computing power, we rely on 1D stellar models. In this work we create our models with \texttt{MESA r11701}. The 1D stellar evolution code \textit{MESA} is open source, and has been continually developed over the last decade \citep{paxton.etal:2011,paxton.etal:2013,paxton.etal:2015,paxton.etal:2018,paxton.etal:2019}. The current implementation of rotation in \texttt{MESA} is described in \cite{paxton.etal:2013} with updates in \cite{paxton.etal:2019}. 

These models are an expansion of the \texttt{MIST} stellar model grid \citep{dotter:2016,choi.etal:2016}, and we adopt the physics mentioned therein. Until now, \texttt{MIST} has used an ad hoc treatment of rotation for stars with convective envelopes, scaling the rotation rate from its full value at stellar masses $>1.8\ \rm \rm M_{\odot}$ towards zero at $1.2\ \rm \rm M_{\odot}$ and below. In addition, models were previously made to rotate only once they reach the zero age main sequence (ZAMS); models are now initialized during the pre-main sequence (PMS) with rotation. The methodology described below has been developed to implement a more realistic treatment of stellar spin down for low mass stars that experience magnetic braking. For this work, we have calculated solar metallicity models, with masses ranging from $0.1-1.3\ \rm M_{\odot}$.

\subsection{Internal Angular Momentum Transport} \label{ssec:jtransport}

\texttt{MESA} uses a diffusive equation \citep{endal.sofia:1978, pinsonneault.etal:1989,heger.langer.woosley:2000} to handle angular momentum transport \citep{paxton.etal:2013}. The equations for rotational chemical mixing (omitted here for brevity, but see \citealt{heger.langer.woosley:2000} for the relevant formulation) are similar. The variation of angular velocity $\omega$ with time within the star takes the form 

\begin{equation} \label{eq:mesa.omdot}
\begin{aligned}
\left(\frac{\partial\omega}{\partial t}\right)_m = \frac{1}{i}\left(\frac{\partial}{\partial m}\right)_t \left[(4\pi r^2 \rho)^2 i \nu \left(\frac{\partial \omega}{\partial m}\right)_t\right] \\ - \frac{2\omega}{r}\left(\frac{\partial r}{\partial t}\right)_m \left(\frac{1}{2}\frac{d \ln i}{d \ln r}\right)
\end{aligned}
\end{equation}
with $i$, $\rho$, $m$, $r$ as the specific moment of inertia, mass density, mass, radius, and $\nu$ as the turbulent viscosity. The turbulent viscosity is the sum of various diffusion coefficients, setting the strength of diffusive transport via e.g., convection and various hydrodynamical mixing processes from rotation induced instabilities. In addition, a constant, arbitrary diffusion coefficient may be set within \texttt{MESA}, representing some additional background mixing source(s), discussed below. 

There is some evidence that additional sources of diffusion may be necessary within this formalism, which could come from internal magnetic fields \citep{eggenberger.maeder.meynet:2005}, or possibly gravity waves \citep{charbonnel.talon:2005} (both of which are physical processes that transport angular momentum in a non-diffusive manner, adding complexity, e.g., see \citealt{rogers:2015,rogers.mcelwaine:2017}), as \cite{denissenkov.etal:2010} and \cite{somers.pinsonneault:2016} mention. Additional sources of diffusion are required to reproduce the rotation profile of the Sun, which appears to be nearly solid body \citep{howe:2009}. This additional source of diffusion is a free parameter (referred to as $\nu_0$ going forward) that we tune to reproduce the solar rotation profile at the solar age for a $1\ \rm \rm M_{\odot}$ model; this extra source operates throughout the entire star, with $\nu=\nu_{\rm hydro}+\nu_0$ in Eq. \ref{eq:mesa.omdot} (where $\nu$ is the sum of all other diffusion coefficients). Due to the efficient transport of angular momentum in convection zones, solid body rotation develops there, while it generally does not within the radiative zone, allowing differential rotation.

We select a constant $\nu_0$ for all of our models, regardless of mass for simplicity (although \citealt{lanzafame.spada:2015,somers.pinsonneault:2016,spada.lanzafame:2020} find evidence of mass dependence). We discuss possible consequences of this choice in Sec.~\ref{sec:discussion}. We calibrate $\nu_0$ to roughly reproduce the differential rotation observed in the Sun, where the core rotates $\lesssim 10\%$ faster than the outer layers (e.g., \citealt{couvidat.etal:2003}). We additionally use the spread in observed $P_{\rm{rot}}$ for $\sim 1\ \rm M_{\odot}$ stars in open clusters ranging from $2$ to $2500$ Myr to calibrate $\nu_0$. Lower $\nu_0$ produces weaker core-envelope coupling, and stronger differential rotation. The degree of differential rotation scales with rotation rate, and stronger differential rotation decreases angular momentum transport efficiency between the core and envelope, which hinders the efficiency of angular momentum loss. Thus, rapidly rotating stars spin down less efficiently than slow rotators, causing a spread in the predicted $P_{\rm{rot}}$ of our models over time. Decreasing $\nu_0$ causes stronger differential rotation, increasing this spread, which does not match open cluster data (e.g., see \ref{f:m15prot}). In this way, cluster data at late ages ($>1$ Gyr) provides a soft lower limit on the value of $\nu_0$ in our models. We adopt a default, solar-calibrated value $\nu_0=2\times10^{4}\ \rm{cm^2}\ \rm{s^{-1}}$ to satisfy both constraints.

The diffusive approximation of angular momentum transport is not used unanimously in 1D stellar models. Codes such as \texttt{GENEC} \citep{eggenberger:2008,ekstrom.etal:2012,georgy.etal:2014} and recent versions of \texttt{STAREVOL} \citep{amard.etal:2016,amard.etal:2019} use an advective-diffusive equation described in \cite{zahn:1992,maeder.zahn:1998} and \cite{mathis.zahn:2004} instead. Both formalisms are valid approximations, but do produce different model behavior (e.g., compare results in \citealt{brandt.huang:2015a} and \citealt{gossage.etal:2018}).

\section{Braking formalisms} \label{sec:blaws}

 An early model for calculating the angular momentum lost by the Sun due to solar wind interactions came from \cite{weber.davis:1967}. It was observed later in \cite{kraft:1967} and \cite{skumanich:1972} that stars spin down as they age. The equatorial surface velocity $v_{\rm eq}$ appeared to be proportional to $t^{-1/2}$ (with $t$ as stellar age); this relation is now often referred to as the Skumanich law. Generally, more slowly rotating low mass stars in open clusters are observed to follow the Skumanich law, although more rapid rotators appear not to. Research has sought a model to describe the physical mechanisms that may cause stars to lose angular momentum as they age, with some early generalizations of the \cite{weber.davis:1967} model coming from \cite{kawaler:1988} and \cite{krishnamurthi.etal:1997}.
 
 \subsection{Initial spin evolution}\label{ssec:ini}

 In understanding the physics of magnetic braking during PMS evolution, there are obfuscating, yet interlinked physical processes that must be addressed to properly model the rotation evolution of stars (\citealt{bouvier.etal:2014} gives an overview). One such process is disk locking. Disk locking is an observationally inferred process \citep{koenigl:1991,shu.etal:1994} of the rotational evolution of these stars on the PMS, where stars appear to transfer angular momentum between their protostellar disk and themselves, such that they maintain a nearly constant rotation rate for some time (the disk locking time). Also see e.g., \cite{matt.pudritz:2005,matt.pudritz:2008a,matt.pudritz:2008b,matt.etal:2012b,gallet.zanni.amard:2019} for more recent models of PMS star-disk interactions. To avoid initializing these models above their critical rotation rates (at which outward centrifugal force is equivalent to surface gravity), rather than starting at an initial angular velocity $\Omega_i$ set by $P_{\rm{rot,i}}$, we initialize the models with a lower angular velocity, and allow them to spin up to their target $P_{\rm{rot,i}}$ through contraction. Once the models have contracted enough, and have reached their target angular velocity (corresponding to $P_{\rm{rot,i}}$), we force the stars to maintain a constant rotation period until the end of their disk locking time. This is similar to the procedures of e.g., \cite{gallet.bouvier:2013} and \cite{amard.etal:2019}. Incorporating a more physically motivated manner of initializing these models by following the developments of e.g., \cite{matt.etal:2012b} or\cite{gallet.zanni.amard:2019} for disk locking would be an important next step, and is a goal of future work.
 
 The disk locking time influences the initial spin evolution of these stars, and is expected to last until about $2-5$ Myr, or so. It is not well understood how this process works in detail, nor specifically how long the disk locking time should be, or what all of the conditions that could affect it might be. Some evidence points towards solar mass stars spinning up (through PMS contraction) between about $2$ and $13$ Myr, where solar mass stars with periods faster than $1$ day do not appear in data for the Orion Nebular Cluster, NGC 6530, NGC 2264, nor NGC 2362 (1, 1.65, 2, and 5 Myr respectively; see e.g., \citealt{gallet.bouvier:2013}). Disks appear to dissipate largely within $13$ Myr, the age of h Per \citep{currie.etal:2007,fedele.etal:2010}. As the physics of disk locking and PMS angular momentum evolution are highly uncertain, we simply select a representative disk locking time of 3 Myr for all models. We did test results with the variable disk locking time scheme used in \cite{amard.etal:2019}, but found it had minimal impact on our results.
 
 With our disk locking time set to 3 Myr, we have used the observed rotation periods for stars in the range $0.95-1.05$ (where colors were converted to masses via our evolution tracks) in the clusters NGC 6530 \citep{henderson.etal:2012} and NGC 2264 \citep{affer.etal:2013} at 2 and 3 Myr respectively, to inform our choice of initial rotation periods. Following the end of disk locking, at 3 Myr in our modelling, models contract during PMS evolution and freely spin up, decreasing their rotation period. We have selected initial rotation periods to match the rotation period range observed in h Per \citep{moraux.etal:2013} at 13 Myr, an age when all stars have largely lost their disks; this is similar to the procedure in \cite{amard.etal:2019}. Stars in h Per with masses estimated as ranging from $0.4$ to $1.4\ \rm \rm M_{\odot}$ generally have rotation periods ranging from about $0.3$ to $10$ days. We adopt an irregularly spaced grid of initial rotation periods ($P_{\rm{rot,i}}$, the initial period at which our disk locked models rotate for ages less than 3 Myr) for our models: 1.5, 3, 4.5, 6, 8, and 12 days, which through PMS spin up, roughly covers the distribution of rotation periods observed by \cite{moraux.etal:2013} at 13 Myr. The most rapidly rotating stars in the \cite{moraux.etal:2013} sample reach up to $P_{\rm{rot,i}}\approx0.2$ days; we have calculated models at $P_{\rm{rot,i}}= 0.4, 0.6$ and 0.8 days, but find that these models near solar mass reach super-critical rotation velocities. Therefore, we do not show our models with $P_{\rm{rot,i}}<1.5$ days in our main results. We do find that these rapid rotators may be needed to reproduce the lithium abundance dispersion, including the low solar value (e.g., as observed in \citealt{carlos.etal:2019}), as we discuss in Sec. \ref{ssec:res.li7}; this issue is discussed further also in Sec. \ref{ssec:disc.fastp}.
 
 Although we adopt a flat distribution of rotation periods here for demonstrative purposes, the detailed distributions of stellar rotation periods seems to be dependent on mass dependent factors. Lower masses possess a more pronounced population of fast rotators than higher masses do \citep{moraux.etal:2013}. Some of the fast rotators are likely the products of binary interaction (e.g. \citealt{douglas.etal:2017}), but e.g., after removing binary candidates, \cite{moraux.etal:2013} found that lower mass stars appear to possess a more significant population of fast rotators compared to their high mass counterparts. This may also be seen at $\sim$10 Myr in the Upper Scorpius (USco) association. \cite{somers.etal:2017} and \cite{rebull.etal:2018} describe a mass-rotation correlation wherein lower mass M dwarfs appear to rotate much faster than higher mass M dwarfs. This could mean, for example, that lower mass M dwarfs exhibit shorter disk locking times so that they spin up sooner than higher mass stars, or that some other mass dependent initial condition exists, which will likely need to be accounted for in models going forward.
 
 In regard to this, rapidly rotating stars with masses $\lesssim 0.6\ \rm \rm M_{\odot}$ have been difficult to understand. Recent models have worked to incorporate additional physics that utilizes data constrained by magnetic field activity, such as \cite{wright.etal:2011,wright.drake:2016,wright.etal:2018}. Two braking models \citep{matt.etal:2015,garraffo.etal:2018} created to model the spin evolution of stars with surface magnetic fields are included in this study and described below.

\subsection{\cite{matt.etal:2015} braking}\label{ssec:m15form}

\cite{matt.etal:2015} produced a model that took advantage of more extensive rotation period data that had become available around the same time \citep{irwin.bouvier:2009,bouvier.etal:2014}. Additionally, \cite{matt.etal:2015} incorporated physically motivated stellar parameter scalings (from the MHD simulations of \citealt{matt.pudritz:2008a,matt.etal:2012}, as also included e.g., by \citealt{gallet.bouvier:2013,vansaders.pinsonneault:2013}) as an update to previous stellar wind angular momentum loss laws. Following \cite{matt.etal:2015} and the implementation described in \cite{amard.etal:2019}, we compute the torque due to magnetic braking as

\begin{equation} \label{eq:m15.jdot.unsat}
 \dot{J} = -\mathcal{T}_0\left(\frac{\tau_c}{\tau_{c,\odot}}\right)^p \left(\frac{\Omega}{\Omega_{\odot}}\right)^{p+1},\ \rm{unsaturated}  
\end{equation}
when the magnetic field is unsaturated, and
\begin{equation} \label{eq:m15.jdot.sat}
 \dot{J} = -\mathcal{T}_0\chi^p \left(\frac{\Omega}{\Omega_{\odot}}\right),\ \rm{saturated}  
\end{equation}
where 
\begin{equation} \label{eq:m15.jdot.const}
 \mathcal{T}_0 = K\left(\frac{R}{R_{\odot}}\right)^{3.1} \left(\frac{M}{\rm M_{\odot}}\right)^{0.5} \gamma^{-2m}    
\end{equation}
and $\gamma = \sqrt{1+(u/0.072)^2}$ from Eq. (8) of \cite{matt.etal:2012}, where  $u$ is the ratio of rotation velocity to critical rotation velocity, $v/v_{\rm{crit}}$. The constants $K$, $m$, $p$, and $\chi$ are free parameters, calibrated to data. The values that we have adopted are collected in Table \ref{tab:m15.params}.

\begin{table}[h!]
    \centering
    \caption{Our adopted parameters in \cite{matt.etal:2015} braking model, with those of \cite{amard.etal:2019} for comparison.}
    \begin{tabular}{||c|c|c||}
         Parameter &  This Work & \cite{amard.etal:2019} \\
         \hline
        K & $1.4\times 10^{30}\ \rm{erg}$ & $7\times 10^{30}\ \rm{erg}$ \\
        m & $0.22$ & $0.22$ \\
        p & $2.6$ & $2.1$ \\
        $\chi$ & $14$ & $14$ \\
    \end{tabular}
    \label{tab:m15.params}
\end{table}

The terms unsaturated and saturated in Eqs. \ref{eq:m15.jdot.unsat} and \ref{eq:m15.jdot.sat} refer to two apparent regimes of stellar magnetic activity. These regimes have been found in e.g., \cite[][and references therein]{wright.etal:2018}, where the X-ray luminosity of stars (a proxy for magnetic activity) appears to correlate strongly with the Rossby number $R_o = (P/\tau_c)$, and becomes nearly constant below some critical value of $\rm R_o$ (i.e., the saturated regime, where $R_o < R_{o_{\rm sat}}$). The parameter $\chi=R_{o_{\odot}}/R_{o_{\rm sat}}$ relies on this critical value, $R_{o_{\rm sat}}$, which was measured in \cite{wright.etal:2018} to be $R_{o_{\rm sat}}=0.14$; the solar $R_{o_{\odot}}$ is believed to be around 2 (e.g., see \citealt{see.etal:2016}), and thus we take $\chi=14$. The constant $m$ we set to $0.22$, adopting the value from e.g., \cite{matt.etal:2015,amard.etal:2019}. The constants $K$ and $p$ have been solar calibrated to reproduce the rotation period of the Sun (we adopt $P_{\odot}=28$ days at the age of the Sun $4.6$ Gyr). For $K$, we adopt a value $1.4\times10^{30}$; for $p$ we adopt $2.6$. The constant $p$ was set to reproduce the spread of rotation rates observed in the Pleiades at $125$ Myr for roughly $1\ \rm M_{\odot}$ stars according to data from \cite{rebull.etal:2018}. In calculating $\rm R_o$, we have adopted a different definition of the convective turnover time ($\tau_c$) than \cite{amard.etal:2019} did in their implementation of \cite{matt.etal:2015}.

Semi-empirical approaches have estimated expected values of $\tau_c$, such as in \cite{wright.etal:2011}, but it is  not clear precisely where $\tau_c$ should be calculated within the interior of a 1D stellar model. \cite{cranmer.saar:2011} derived an effective temperature dependent function to predict $\tau_c$, but this relationship may not be universally applicable to all stars. A number of stellar models base their convective turnover times on mixing length theory \citep{bohm-vitense:1958}, placing the calculation at some multiple (often $0.5$ times) of the pressure scale height, as defined in \cite{gilliland:1985}. This roughly places the calculation within range of a convective eddy assumed to be related to the dynamo generating the magnetic field, tying the definition of $\tau_c$ loosely to the magnetic dynamo. However, it is also unclear if fully convective stars generate surface magnetic fields with the same dynamo mechanisms that partially convective stars use \citep{mullan.macdonald:2001,reiners.basri:2007,irwin.etal:2011,wright.drake:2016,wright.etal:2018}. See \cite{charbonnel.etal:2017} for a recent analysis of various definitions of the convective turnover time in relation to data. 

We calculate $\tau_c$ as
\begin{equation} \label{eq:tauc}
 \tau_c(r) = \alpha_{\rm MLT} H_P(r)/v_c(r)
\end{equation}
where $\alpha_{\rm MLT}$ is the convective mixing length parameter ($\alpha_{\rm MLT}=1.82$ in our models; \citealt{choi.etal:2016}), $H_P(r)$ is the scale height, and $v_c(r)$ is the convective velocity within the star at radius $r$. We calculate $\tau_c$ one half a pressure scale height above the bottom of the outermost convection zone, defined as where $r=r_{\rm BCZ}+0.5 H_P(r)$, and where $r_{\rm BCZ}$ is the position of the bottom of the outer convection zone. This is a simple approach that provides better agreement to the semi-empirical trend of $\tau_c$ with mass in \cite{wright.etal:2011} at an age of 1 Gyr (as assumed in that work) for our models than the definition of $\tau_c$ at one half of a pressure scale height in \cite{gilliland:1985} does.

\subsection{\cite{garraffo.etal:2018} braking}

We have also included the braking model of \cite{garraffo.etal:2018}. This model is based on MHD simulations \citep{reville.etal:2015,garraffo.drake.cohen:2015,garraffo.drake.cohen:2016} that suggest the efficiency of magnetic braking depends on the complexity of the stellar magnetic field, not just its strength. The model combines an expression for angular momentum loss that is based on the Skumanich law, designated $\dot{J}_{\rm dip}$, with a function $Q_{J}(n)$ to modulate this basic loss formalism by the magnetic field complexity, parameterized as $n$ \citep{garraffo.drake.cohen:2016}. The full angular momentum loss formalism is 
\begin{equation} \label{eq:g18.jdot}
 \dot{J} = \dot{J}_{\rm dip} Q_{J}(n)   
\end{equation}
with
\begin{equation} \label{eq:g18.jdip}
 \dot{J}_{\rm dip} = c \Omega^3 \tau_{\rm c}   
\end{equation}
representing Skumanich spin down ($P\propto t^{-1/2}$) that the model converges to when the magnetic field becomes dipolar. Here, $c$ is a free parameter setting the overall strength of $\dot{J}_{\rm dip}$, $\Omega$ is the angular velocity of the stellar surface, and $\tau_{\rm c}$ is the convective turnover time.


The function $Q_J (n)$ was derived in \cite{garraffo.drake.cohen:2016} and serves to modulate the dipolar Skumanich angular momentum loss via magnetic field complexity. It takes the form
\begin{equation} \label{eq:g18.qj}
 Q_{J}(n) = 4.05 e^{-1.4n} + \frac{n-1}{60 B n}
\end{equation}
where $B$ is the magnetic field strength, and $n$ parameterizes the magnetic field complexity. In practice, this second term only begins to matter in relatively complex magnetic fields where $n>7$. Instead of following \cite{garraffo.etal:2018}, who imposed an upper limit at $n=7$ so that this term may be ignored, and $Q_J (n)$ takes the form
\begin{equation} \label{eq:g18.qjsimple}
 Q_{J}(n) = 4.05 e^{-1.4n},
\end{equation}
we simply ignore the term, and do not place an upper limit on $n$. We do this for simplicity because we currently do not calculate magnetic field strength in our models. The ignored term adds a constant to $Q_J (n)$, such that more complex magnetic fields are even more suppressed than they would be were the term included. We find that our models that reach $n>7$ are often too suppressed regardless of the inclusion of this term, and so capping $n$ has little consequence on our results.

The magnetic complexity number $n$ is itself a function of $\rm R_o$, with $P$ being the rotation period of the star, $P = 2\pi/\Omega$,
\begin{equation} \label{eq:g18.n}
 n = \frac{a}{R_o} + 1 + b R_o   
\end{equation}
representing the magnetic field complexity. In Eq. \ref{eq:g18.n}, $n=1$ corresponds to a dipole field, while higher $n$ represents more complexity and higher order multipoles. Also in Eq. \ref{eq:g18.n}, the free parameters $a$ and $b$ are free parameters. This equation was constructed to match expected magnetic field complexity trends based on Zeeman-Doppler-Imaging \citep{donati.landstreet:2009,marsden.etal:2011,waite.etal:2015,alvarado-gomez.etal:2015}, and \textit{Kepler} gyrochronological observations of stars at $\rm R_o$ greater than about $1-2$ \citep{vansaders.etal:2016,vansaders.pinsonneault.barbieri:2019}. 

\begin{figure}[h!]
    \centering
    \includegraphics[width=0.4\textwidth]{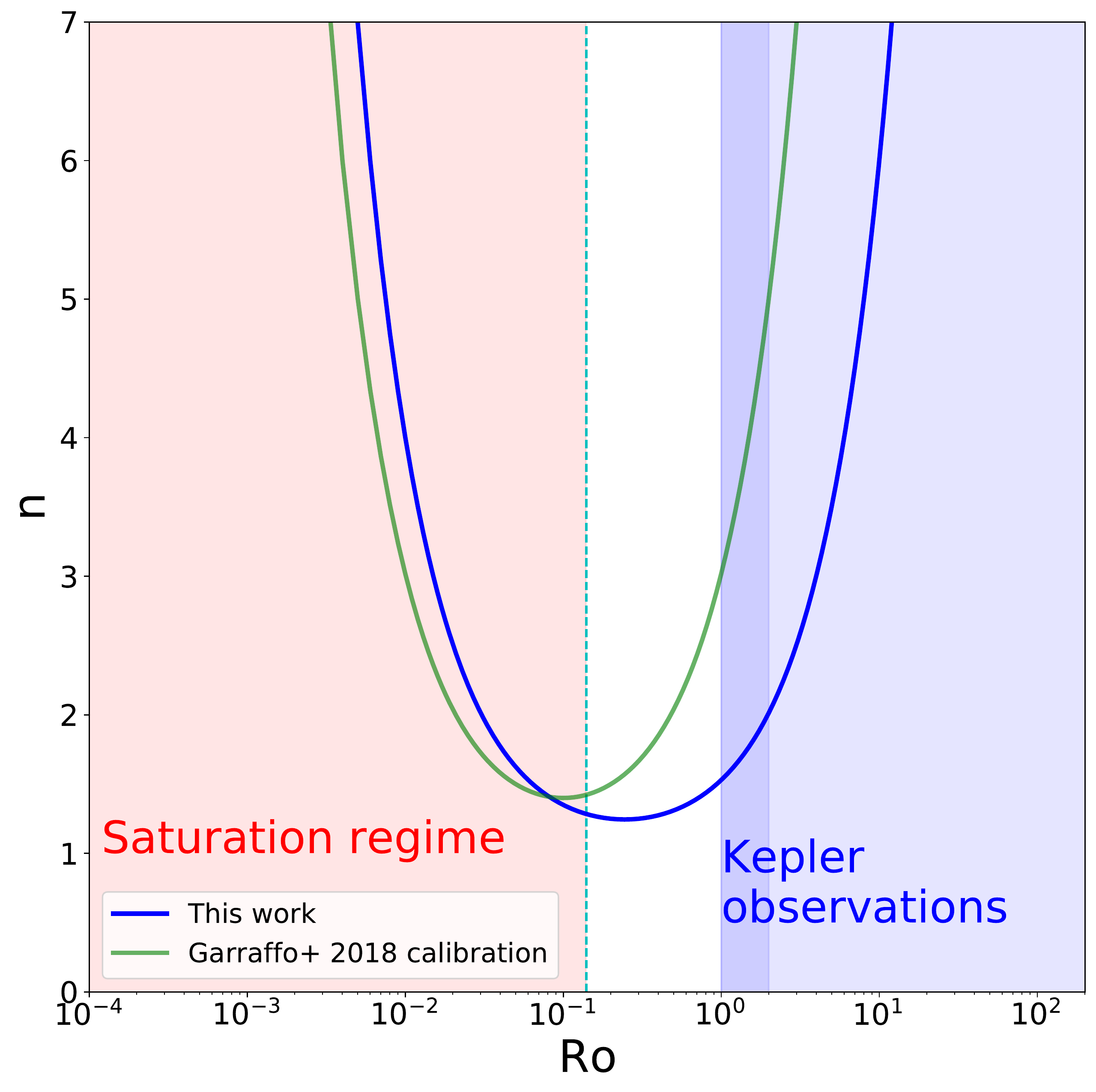}
    \caption{The variation of $n$ (see Eq. \ref{eq:g18.n} and \citealt{garraffo.etal:2018}), the magnetic complexity parameter, with Rossby number, $\rm R_o$. The blue (\textit{Kepler} observations) and red (magnetically saturated) shaded regions are similar to the regions shown in the corresponding figure of \cite{garraffo.etal:2018}, but our adopted saturation point lies at $R_{\rm{o,sat}}=0.14$, rather than their $0.11$, as described in the text.}
    \label{f:g18n}
\end{figure}

The parameters we adopt for this model were arrived at through calibration to cluster data, and are listed in Table \ref{tab:g18.params}. The parameter $c$ controls the overall strength of angular momentum loss in the model, and was tuned to reproduce the solar rotation rate. We arrived at a larger value of $a$ than in \cite{garraffo.etal:2018}. In Eq. \ref{eq:g18.n}, lower values of $a$ mean that $n$ reaches its minimum (becomes dipolar) at lower values of $Ro$; higher values of $a$ mean that it takes higher $\rm R_o$ before a dipolar field is achieved, effectively meaning that models take longer to reach a Skumanich type spin down, and remain spinning faster, longer. Lastly, we changed $b$ to control magnetic suppression at higher $\rm R_o$ (i.e., slower rotation and typically older ages). We adopt a smaller value of $b$ so that our solar model reaches the solar rotation rate as well as matches the rotation periods of solar-like stars observed in open cluster data prior to $4.6$ billion years. A larger $b$ means that at later ages, stars regain magnetic complexity, and a smaller $b$ means that stars stay mostly dipolar (and more closely follow Skumanich spin down) at later ages. Our adopted values are tuned mostly to reproduce solar rotation rates, but it is not clear that $n$ should be universally assigned as in Eq. \ref{eq:g18.n} for all stellar masses.

\begin{table}[h!]
    \centering
    \caption{Our adopted parameters in the \cite{garraffo.etal:2018} braking model.}
    \begin{tabular}{||c|c|c||}
         Parameter &  This Work & \cite{garraffo.etal:2018} \\
         \hline
        a & $0.03$ & $0.02$ \\
        b & $0.5$ & $2.0$ \\
        c & $3\times 10^{41}\ \rm{g}\ \rm{cm^{-2}}$ & $1\times 10^{41}\ \rm{g}\ \rm{cm^{-2}}$ \\
    \end{tabular}
    \label{tab:g18.params}
\end{table}

As $n$ (magnetic field complexity) increases, the portion of the stellar surface covered by open field lines decreases. Fewer open field lines leads to less outgoing stellar wind as complexity increases, reducing the capacity for magnetic braking to act in slowing the star down. Thus, more complex stars will tend to rotate faster, and less complex stars more slowly. The fast and slow branch rotators observed in open clusters may be explained through this relationship between field complexity and braking efficiency. Higher magnetic field complexity leads to greater suppression of the braking process. This has been shown previously in \cite{garraffo.etal:2018}, though models in that case were not evolved with these effects in place; rather, they were applied on pre-computed stellar models. Here, we self-consistently evolve the models with this braking scheme.  

\begin{figure*}[ht!]
    \centering
    \includegraphics[width=\textwidth]{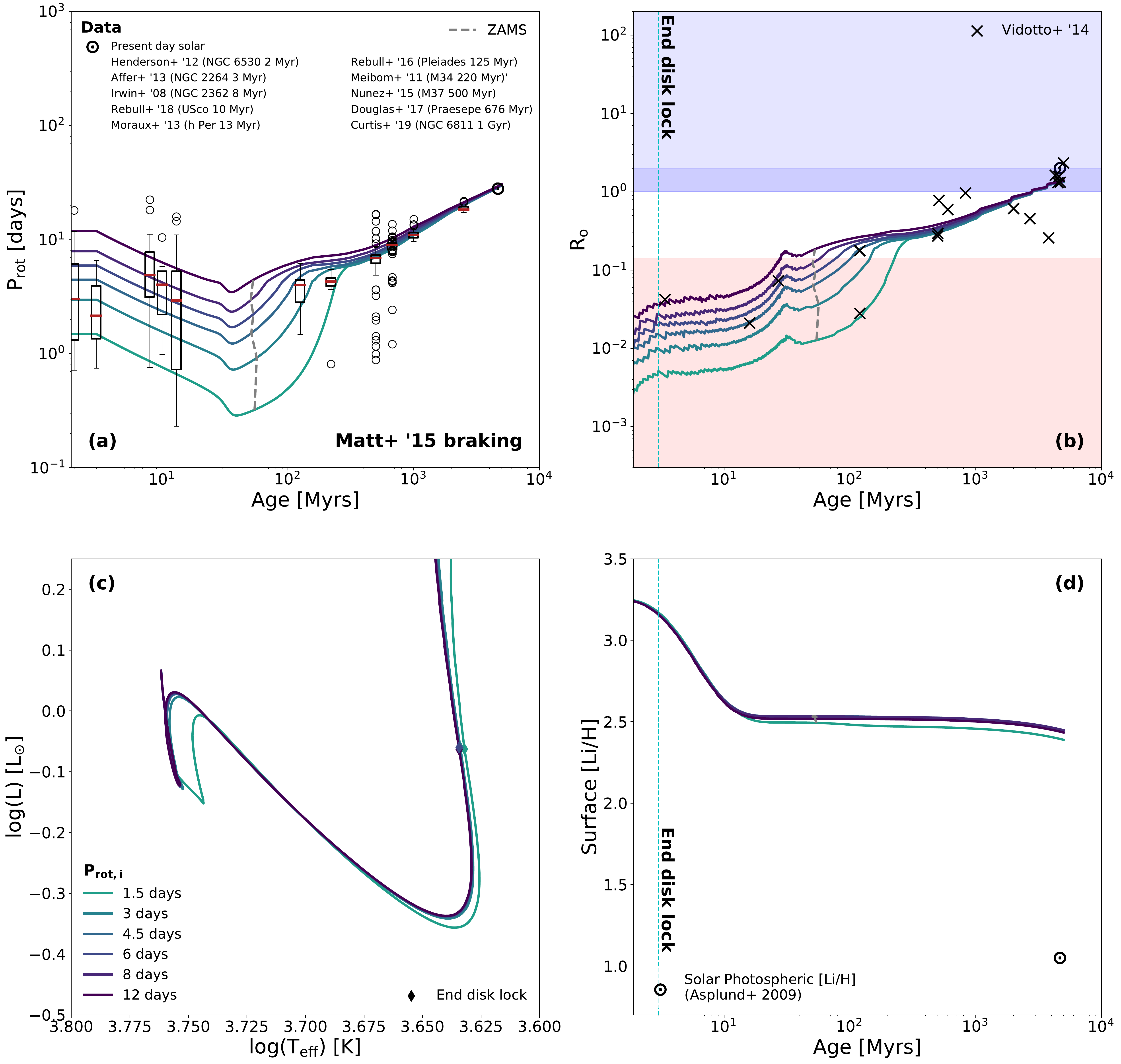}
    \caption{The evolution of a $1 \rm M_{\odot}$, solar metallicity model with our implementation of the \cite{matt.etal:2015} braking model. Shown are the evolution of (a) $\rm P_{rot}$, (b) Rossby number ($\rm R_o$) (c) luminosity and temperature (HRD) from 100 years to 5 Gyr, and (d) surface lithium abundance ($\rm log(^7Li/H) + 12$) with stellar age. Colors for each solid line correspond to initial rotation period (set at 3 Myr), $P_{\rm{rot,i}}=1.5,3,4.5,6,8$, and 12 days are shown, progressing from blue-green to dark blue progressively. Rotation periods for stars with $0.95\ \rm M_{\odot}\leq M \leq 1.05 \rm M_{\odot}$ are shown as a box-and-whiskers plot in panel (a); outliers are shown as empty circles and are likely binaries \citep{douglas.etal:2017}. Crosses in panel (b) show $\rm R_o$ values from \cite{vidotto.etal:2014} for the same mass range. The grey dashed line in panels (a), (b), (c) marks ZAMS; the cyan vertical line in panels (c) and (d) marks the end of disk locking (3 Myr), while colored diamonds represent this in panel (c).
    \label{f:m15prot}}
\end{figure*}

\section{Results} \label{sec:results}

\begin{figure*}[ht!]
    \centering
   \includegraphics[width=\textwidth]{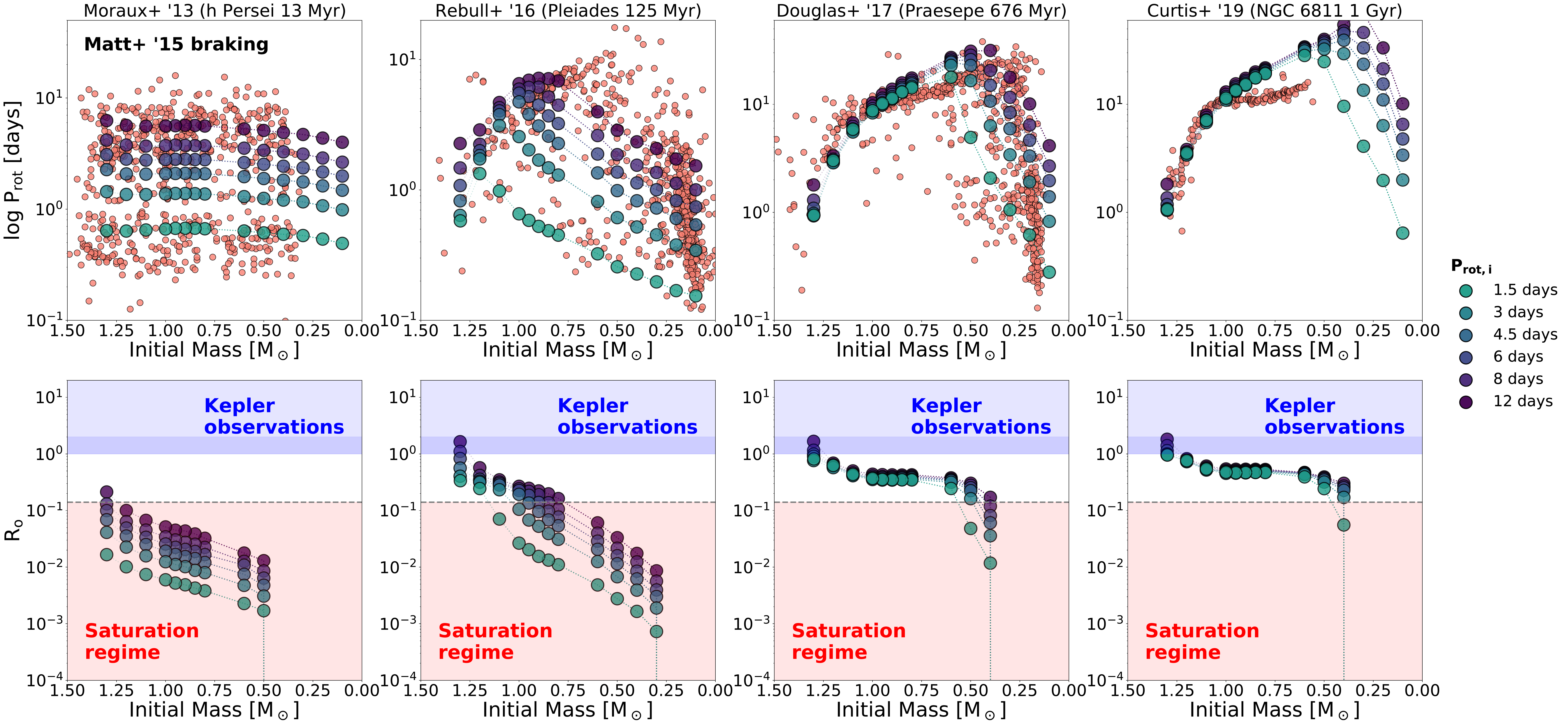}
    \caption{Our models adopt the \cite{matt.etal:2015} braking model in this figure. \textbf{Top row:} observed rotation period $\rm P_{rot}$ is plotted vs. stellar mass for h Persei (13 Myr), the Pleiades (125 Myr), the Praesepe (676 Myr), and NGC 6811 (1 Gyr) in the top row. Data are plotted as salmon colored dots from \cite{moraux.etal:2013,rebull.etal:2016,douglas.etal:2017,curtis.etal:2019}, respectively. Our stellar models are overplotted as colored dots. Colors for the dots are the same as described in Fig. \ref{f:m15prot}. \textbf{Bottom row:} Rossby number ($\rm R_o$) vs. age at ages corresponding to the top row. The blue (\textit{Kepler} observations) and red (magnetically saturated) shaded regions correspond to the same regions as in Fig. \ref{f:g18n} for reference.}
    \label{f:m15_protevo}
\end{figure*}

\begin{figure*}[ht!]
    \centering
    \includegraphics[width=\textwidth]{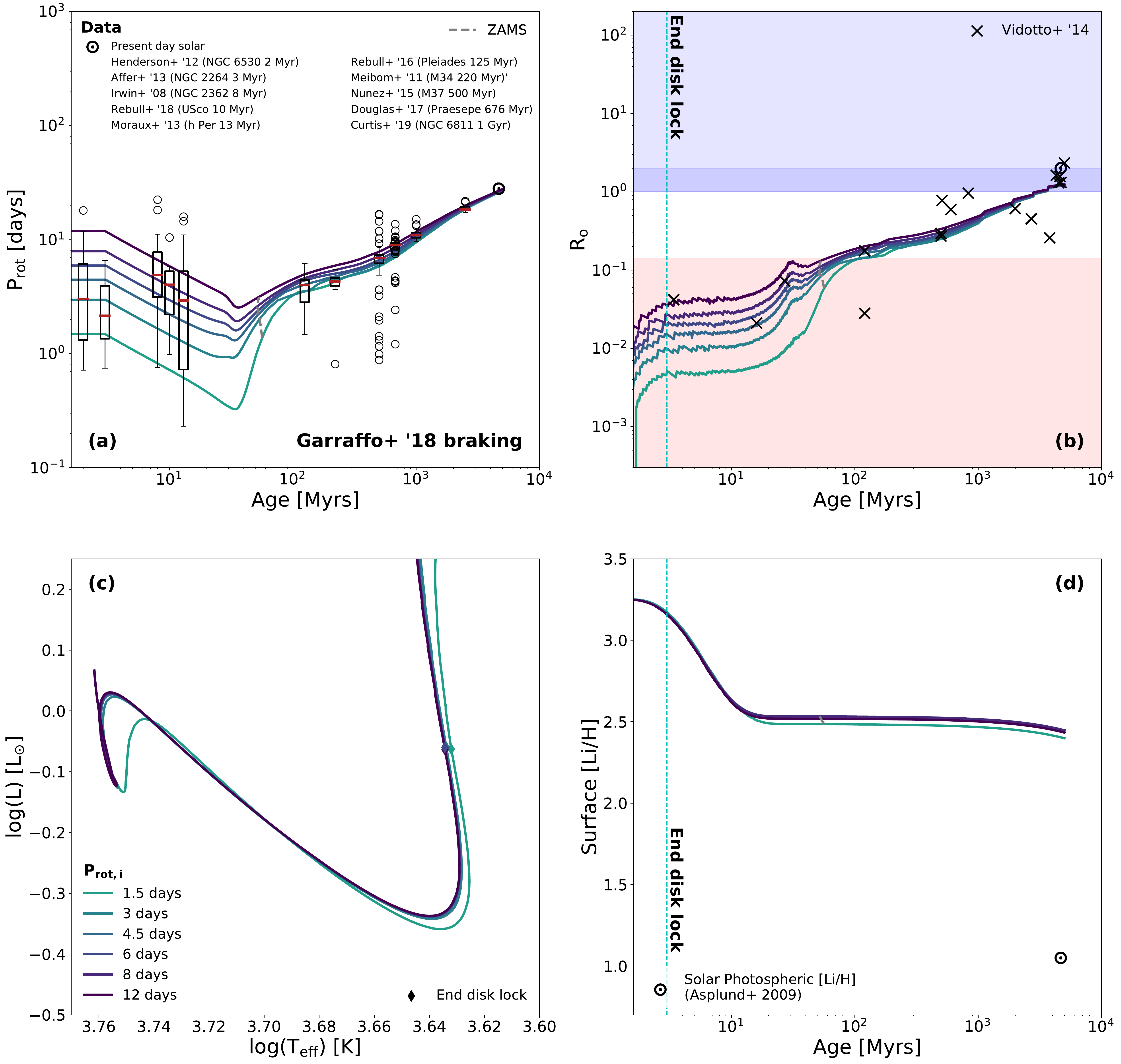}
    \caption{The same as Fig. \ref{f:m15prot}, except with our implementation of the \citep{garraffo.etal:2018} braking model. The colored lines again correspond to our models, similar to what is described in the caption of Fig. \ref{f:m15prot}.}
    \label{f:g18prot}
\end{figure*}

\begin{figure*}[ht!]
    \centering
   \includegraphics[width=\textwidth]{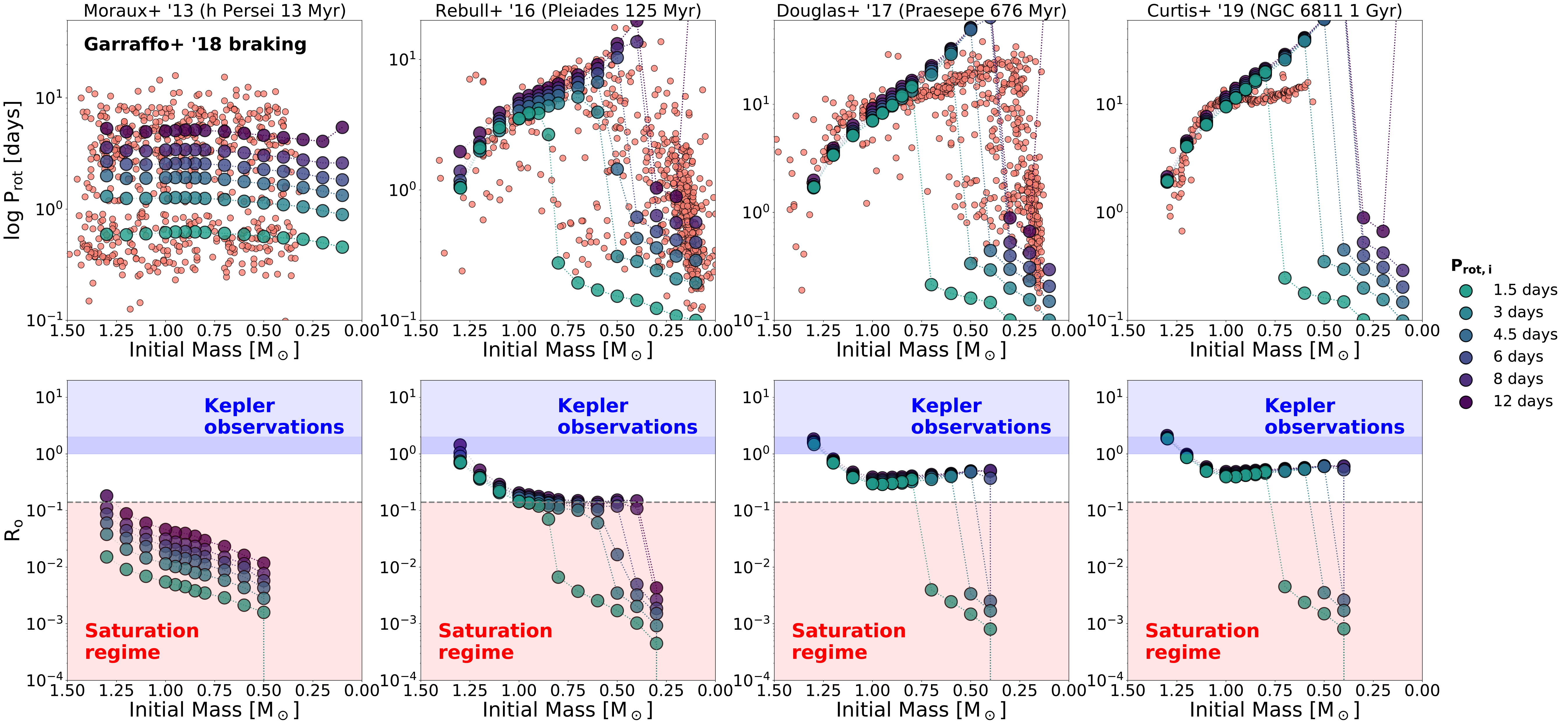}
    \caption{The same as Fig. \ref{f:m15_protevo}, except our models adopt the \cite{garraffo.etal:2018} braking model here. Note that the bimodal nature of rotation period data in e.g., the Pleiades and the Praesepe is reproduced; the model shows a tight sequence of slow rotators at higher masses (although often too slow), and jumps to faster rotation rates at lower masses, sparsely populating the region in between these regimes. Compare this to how the \cite{matt.etal:2015} model produces many more rapid rotators at low masses that do not reflect the observed scarcity of such stars in Fig. \ref{f:m15_protevo}, top row.}
    \label{f:g18_protevo}
\end{figure*}

In this section we present the evolution of rotation period with time according to our models when using the \cite{matt.etal:2015} (Sec. \ref{ssec:res.m15}), or the \cite{garraffo.etal:2018} braking model (Sec. \ref{ssec:res.g18}). The results are displayed respectively in Figs. \ref{f:m15prot} and \ref{f:m15_protevo} for the \cite{matt.etal:2015} braking results, while Figs. \ref{f:g18prot} and \ref{f:g18_protevo} pertains to the same for the \cite{garraffo.etal:2018} braking model. The ages of open clusters included in the comparisons range from around $3$ to $2500$ Myr. The clusters shown are NGC 6530 ($\sim 2$ Myr old; \citealt{henderson.etal:2012}), NGC 2264 ($\sim 3$ Myr old; \citealt{affer.etal:2013}), NGC 2362 ($\sim 8$ Myr old; \citealt{irwin.etal:2008}), USco ($\sim 10$ Myr; \citealt{rebull.etal:2018}), h Persei (h Per, $\sim 13$ Myr old; \citealt{moraux.etal:2013}), the Pleiades ($\sim$125 Myr old; \citealt{rebull.etal:2016}), the Praesepe ($\sim$676 Myr old; \citealt{douglas.etal:2017}), and NGC 6811 ($\sim$1 Gyr old; \citealt{curtis.etal:2019}). We have also utilized M34 ($\sim$220 Myr old; \citealt{meibom.etal:2011}), M37 ($\sim$500 Myr old; \citealt{nunez.etal:2015}), and NGC 6819 ($\sim$2.5 Gyr old; \citealt{meibom.etal:2015}) data for comparisons, but these are not all shown in Figs.~\ref{f:m15_protevo} and \ref{f:g18_protevo} for the sake of space, and their ages do not vary largely from clusters that have been included. The solar mass stars from these data sets are shown in Figs. \ref{f:m15prot} and \ref{f:g18prot}, however. We reiterate that lower mass M dwarfs are observed to contain a larger fraction of rapid rotators, than higher mass stars do (e.g., \citealt{somers.etal:2017,rebull.etal:2018}). In this study, we have not attempted to match the observed distribution of rotation rates in such detail, rather we intend to display simply a representative range of observed rotation rates; matching distributions is a goal of future work. In
Sec. \ref{ssec:res.hrd} we discuss model behavior on the Hertzsprung-Russell diagram (HRD), and in Sec. \ref{ssec:res.li7} the surface lithium depletion seen in our solar mass models (panels (c) and (d), respectively in Figs. \ref{f:m15prot} and \ref{f:g18prot}). In Figs. \ref{f:m15prot} and \ref{f:g18prot}, the data that we compare to is near solar mass, in the range $0.95$ to $\rm 1.05\ M_{\odot}$ (based either directly on masses or on colors provided in the data tables; colors were converted to masses via our isochrones in the latter case).

In interpreting our results, it is important to bear in mind that the angular momentum evolution of our models (and thus our results in general) is still linked to the diffusion parameter $\nu_0$ (Sec. \ref{ssec:jtransport}), representing the degree of core-envelope coupling. Core-envelope coupling affects the angular momentum transfer timescale, and can produce different rotation period evolution patterns. We do not perform an extensive parameter study of $\nu_0$ in this work, but rather adopt a single value, calibrated to reproduce the differential rotation of the Sun (see e.g., \citealt{pinsonneault.etal:1989,krishnamurthi.etal:1997,denissenkov.etal:2010,somers.pinsonneault:2016} for studies more focused on exploring this). We further note issues related to core-envelope coupling, and their impact on our results in the following subsections.

Overall, our results are in line with those of similar studies, and the original papers of \cite{matt.etal:2015} and \cite{garraffo.etal:2018} in matching open cluster rotation periods (i.e., Figs. \ref{f:m15prot} and \ref{f:m15prot}). Our results for \cite{matt.etal:2015} models qualitatively match those shown in Fig. 7 of \cite{amard.etal:2019}, and our results from the \cite{garraffo.etal:2018} model match the results in Fig. 4 of that respective paper (although \cite{garraffo.etal:2018} performed population synthesis, so plot stellar densities, rather than simply positions on period-mass, or period-color diagrams). All results generically show too much angular momentum loss for sub-solar mass models at 1 Gyr and older ages, and carry similar morphologies to the results shown here in period-mass space, discussed further below.

\subsection{\cite{matt.etal:2015} comparisons}\label{ssec:res.m15}
Results from our models employing the \cite{matt.etal:2015} braking scheme show a plausible reproduction of the cluster data. Agreement is generally good between our $1\ \rm M_{\odot}$ model in panel (a) of Fig. \ref{f:m15prot} and cluster observations, given the presence of outliers (indicated by black crosses). In panel (b), we have included measured Rossby numbers for solar-like stars (sub-sampling the data from \citealt{vidotto.etal:2014} in the mass range $0.95\ \rm M_{\odot}\leq M \leq 1.05 \rm M_{\odot}$), and our calculated Rossby numbers agree fairly well over time.

For the evolution of rotation period regarding all masses, view Fig. \ref{f:m15_protevo}. E.g., for the Pleiades (assuming an age of $125$ Myr) by \cite{rebull.etal:2016}, the overall morphology in $P_{\rm{rot}}$-mass space agrees fairly well, but appears to overestimate the number of stars that rotate quickly, as the data shows a more collapsed sequence of slow rotators by this age. Many lower mass ($\leq 0.6\ \rm M_{\odot}$) stars are just halting contraction and entering the ZAMS around this age and have spun up since the end of disk locking at $3$ Myr. Higher mass stars have already completed contraction prior to this and begun to spin down under the influence of magnetic braking, converging towards the slow (Skumanich) sequence of stars. This model underestimates the slowly rotating stars around $0.5\ \rm M_{\odot}$, and generally overestimates the number of rapid rotators with masses $\lesssim0.8\ \rm M_{\odot}$. 

At the age of the Praesepe (assumed here to be 676 Myr), we compare to data from \cite{douglas.etal:2017}. Our models implementing the \cite{matt.etal:2015} model allow for excellent reproduction of the slow rotators across the entire mass range. The lower mass, fast rotators are reproduced as well. The behavior is similar to what is shown in \cite{matt.etal:2015}, but \cite{douglas.etal:2017} found that their \cite{matt.etal:2015} models predicted too many quickly rotating stars with mass $\leq 0.8\ \rm M_{\odot}$. Our models show similar behavior, with stars in this mass regime rotating fast. Essentially all of the stars $0.3\ \rm M_{\odot}\leq M\leq 0.8\ \rm M_{\odot}$ rotate too quickly, in comparison to the data which shows the majority of stars with these masses are slow rotators. 

Faster rotation rates and lower masses mean shorter $P_{\rm rot}$ and larger $\tau_c$ (because of deeper convection zones at low mass), and thus smaller $\rm R_o$ for these stars. The smaller $\rm R_o$ keeps their magnetic fields in the saturated regime longer, delaying their spin down. This may be seen in Fig. \ref{f:m15prot}, panel(b) for our $1\ \rm M_{\odot}$ model, and in Fig. \ref{f:m15_protevo}, bottom row for all masses. So, while our implementation of \cite{matt.etal:2015} is able to reproduce quickly rotating low mass stars, as observed in the data, it seems to overestimate the number of these stars, especially in the range $0.3\ \rm M_{\odot}\leq M\leq 0.8\ \rm M_{\odot}$, or so.

Towards later ages, data from \cite{curtis.etal:2019} of the open cluster NGC 6811 (estimated at 1 Gyr old) shows that stars have largely converged to a single sequence by this time. \cite{meibom.etal:2015} also observed this in the 2.5 Gyr old NGC 6819 (not shown in Fig. \ref{f:m15prot}). Using \cite{matt.etal:2015}, rotation periods tend to be too large for masses $< 1\ \rm M_{\odot}$. The models correctly predict a collapsed sequence of slow rotators by this age, but have generally experienced braking that is too efficient. Masses $\geq 1\ \rm M_{\odot}$ are predicted fairly well. This is a generic problem in our models (regardless of the braking model), and may be related to missing mass dependency related to core-envelope coupling (we discuss this further in Sec. \ref{ssec:disc.1gyr}; see also  \citealt{lanzafame.spada:2015,somers.pinsonneault:2016,spada.lanzafame:2020} for further research specific to this issue).

\subsection{\cite{garraffo.etal:2018} comparisons}\label{ssec:res.g18}

In Fig. \ref{f:g18prot}, we show an overview of behavior for our solar mass model. The model does well matching the evolution of $P_{\rm{rot}}$ up to 4.6 Gyr, as seen in panel (a). Likewise, it does a fair job matching the data of \cite{vidotto.etal:2014} for measured Rossby numbers of solar-like stars (their data sub-sampled to be in the mass range $0.95\ \rm M_{\odot}\leq M \leq 1.05 \rm M_{\odot}$ here).

Refer to Fig. \ref{f:g18_protevo}, top row, to see the evolution of $P_{\rm{rot}}$ at selected ages for all masses. Using the \cite{garraffo.etal:2018} braking scheme, models are able to reproduce the general morphology of $P_{\rm rot}$-mass space across time, but predict braking that is often too efficient at later ages for masses $0.3\ \rm M_{\odot}\leq M\leq 0.8\ \rm M_{\odot}$ (see Fig. \ref{f:g18_protevo}). At $125$ Myr, the slowly rotating stars near $0.5\ \rm M_{\odot}$ are captured better than when using \cite{matt.etal:2015} braking. Slower models with roughly $\rm{M} > 0.5\ \rm M_{\odot}$ have large enough $\rm R_o$ to enter a regime where $n$ is closer to a dipole at this age (see Fig. \ref{f:g18_protevo} and Fig. \ref{f:g18n}), causing them to spin down more aggressively. Apparently the spin down is sufficient in the case for masses near $0.5\ \rm M_{\odot}$, allowing them to reach periods closer to what is observed at this age.

The implementation of \cite{garraffo.etal:2018} in our models qualitatively matches the morphology observed in the Praesepe, however it tends to predict too much spin down here as well with stars in the mass range $0.3\ \rm M_{\odot}\leq M\leq 0.8\ \rm M_{\odot}$. Still, this model does well in reproducing the quickly rotating stars with masses $< 0.3\ \rm M_{\odot}$, and predicts a collapsed sequence of slow rotators by this age, (albeit rotating too slowly), as well as a sharp drop towards fast rotation with fully convective stars ($\rm{M}<0.3\ \rm M_{\odot}$). Simultaneously capturing these low mass fast rotators, and a tight sequence of slowly rotating, higher mass stars allows this braking model to capture the bimodality observed e.g., in the Pleiades and the Praesepe, which is a relative strength compared to using our implementation of the \cite{matt.etal:2015} formalism. In Fig. \ref{f:g18_protevo}, notice the sharp transition from slow to fast rotation going towards lower masses, rather than the gradual, more continuous behavior seen with the \cite{matt.etal:2015} braking model.

The reason models are able to reproduce the low mass fast rotators and tight sequence of slow rotators simultaneously under this braking model comes down to the parabolic dependency of the magnetic complexity ($n$, Eq. \ref{eq:g18.n}) on $\rm R_o$. In general our low mass models have deeper convection zones and higher $\tau_c$, driving smaller $\rm R_o$ (see Figs. \ref{f:m15_protevo} and \ref{f:g18_protevo}, bottom rows), and faster rotation also decreases $\rm R_o$ further. Due to their low $\rm R_o$ (which may be seen in both Fig. \ref{f:g18_protevo} and Fig. \ref{f:g18prot}), low mass models experience greater magnetic complexity (as per Eq. \ref{eq:g18.n}), suppressing their spin down until later ages. As seen in Fig. \ref{eq:g18.n}, there is a fairly narrow range of $R_o$ over which stars experience dipolar (strong) magnetic braking; higher mass models reach this threshold sooner than lower masses. The higher mass models (with generally higher $\rm R_o$) spin down sooner, and re-enter a phase of increased magnetic field complexity (high $n$) as their $\rm R_o$ increases, slowing their braking once again, and keeping them in a tight sequence of slow rotators.  

 At the age of NGC 6811 ($1$ Gyr), models have spun down too much, similar to results from our implementation of \cite{matt.etal:2015} braking, although the discrepancy is even more severe in this case. Again, models with mass $\geq 1\ \rm M_{\odot}$ are represented fairly well, and the models correctly predict a collapsed sequence of slow rotators. A portion of masses below $0.6\ \rm M_{\odot}$ are predicted to remain quickly rotating, but the observations do not include stars in this mass range for comparison. This suggests that for masses $<1\ \rm M_{\odot}$, there may be some additional mass dependence missing from this braking model, where they should be locked (by high $n$) at smaller $\rm R_o$ than higher masses are (effectively, $b$ in Eq. \ref{f:g18n} would need to increase towards lower masses). However, and as previously mentioned in Sec. \ref{ssec:res.m15}, this issue may be related to mass dependency in core-envelope coupling timescales, as suggested by \cite{spada.lanzafame:2020} (see Sec. \ref{ssec:disc.1gyr} for more on this).

\subsection{Effects on the Hertzsprung-Russell Diagram} \label{ssec:res.hrd}

In Figs. \ref{f:m15prot} and \ref{f:g18prot}, panel (c) shows the evolution of our $1\ \rm M_{\odot}$ model on the HRD. There is a minimal effect going towards lower masses, and the effects are similar for masses up to at least $1.3\ \rm M_{\odot}$ included in this study. The inclusion of PMS rotation causes fast rotators to reach the ZAMS at a cooler temperature than slower rotators. This is slightly more pronounced in the case of \cite{matt.etal:2015} braking, as those models take longer to spin down, and so are rotating faster near ZAMS than models do under our modeling with \cite{garraffo.etal:2018} braking. Faster rotation rates cause stronger centrifugal force on the star, enhancing the effect of gravity darkening (e.g., \citealt{vonzeipel:1924,espinosalara.rieutord:2011,paxton.etal:2019}), and causing stars that rotate fast enough to appear cooler upon reaching the ZAMS. This would suggest that low mass, rotating stars have some spread due to gravity darkening on the MS. Although, this effect is barely noticeable, except with the fastest rotation rates ($P_{\rm{rot,i}\leq 1.5}$ days).

\subsection{Lithium burning} \label{ssec:res.li7}

Additional mixing from stellar rotation has been proposed as a possible means of explaining the surface lithium abundances of $\sim 1\ \rm M_{\odot}$ stars. Observations of stars in open clusters (e.g., \citealt{sestito.randich:2005}), and our own Sun, reveal that solar mass stars appear to have a dispersion in their surface lithium abundances, hinting that something must cause more efficient lithium depletion in otherwise similar stars (see recent solar twin studies in, e.g., \citealt{thevenin.etal:2017,carlos.etal:2019}). \cite{somers.pinsonneault:2016} provides an overview of this, and also shows that including an additional angular momentum transport (parameterized by $\nu_0$ here; refer to Sec. \ref{ssec:jtransport}) may explain the dispersion. Rotating stellar models typically are not able to reproduce the solar lithium depletion via rotational mixing alone (e.g, \citealt{amard.etal:2016,amard.etal:2019}). As seen in Figs. \ref{f:m15prot} and \ref{f:g18prot}, our default models do not reproduce the solar lithium abundance. We discuss this further in Sec. \ref{ssec:disc.li7}.

\section{Discussion} \label{sec:discussion}

Our implementations of \cite{matt.etal:2015} and \cite{garraffo.etal:2018} allow us to compare a more traditional magnetic braking model with one that models the effect of magnetic complexity on stellar spin down. According to our results, magnetic complexity may be valuable in providing a mechanism for periods of reduced magnetic braking as stars evolve. This may aid in reproducing the solar lithium abundance, and in matching the behavior of rapidly rotating low mass stars at ages of the Praesepe and prior, and as a possible mechanism for the observed stall in angular momentum loss \citep{agueros.etal:2018,curtis.etal:2019} at later ages.

\subsection{Discrepancies near 600 Myr}\label{disc.600myr}

At the age of the Praesepe (assumed $676$ Myr in this study), spin down via our implementation of \cite{matt.etal:2015} appears to overestimate the presence of rapid rotators with $0.3\ \rm M_{\odot}<M<0.8\ \rm M_{\odot}$. Meanwhile, our models based on \cite{garraffo.etal:2018} predict a more collapsed sequence of stars in this mass regime, but they tend to spin too slowly. In both cases, it is possible that re-calibrations of, or modifications to, the braking model may be necessary in this mass regime. Angular momentum loss appears to be too efficient in the case of \cite{garraffo.etal:2018} braking, and too weak in the case of \cite{matt.etal:2015}.

\subsection{Overestimated spin down after 1 Gyr}\label{ssec:disc.1gyr}
Our models generally experience too much angular momentum loss by 1 Gyr; compared to observations, the rotation periods of $\rm{M}<1\ \rm M_{\odot}$ models are too long. \cite{curtis.etal:2019} discuss that at 1 Gyr, there appears to be an epoch of stalled angular momentum loss in stars $\leq 1\ \rm M_{\odot}$. Angular momentum loss seems to become less efficient around the age of the Prasepe for some time, as stars at the later age of NGC 6811 do not appear to have slowed down by very much more. \cite{curtis.etal:2019} find that gyrochronology models tend to predict too much angular momentum loss, as our models do. The root of this issue is unclear, but may be due to metallicity effects \citep{angus.etal:2015}, or perhaps a mass dependence on the angular momentum loss rate that is missing from current models (see also e.g., \citealt{agueros.etal:2018}).

One explanation for this discrepancy could lie in the parameter $\nu_0$ (Sec. \ref{ssec:jtransport}), which is related to the degree of core-envelope coupling within stars. Stronger coupling causes angular momentum transport to behave closer to a solid body (more efficient core-envelope transport). \cite{somers.pinsonneault:2016} studied this, and found evidence that the level of core-envelope coupling in stars may be mass dependent. \cite{lanzafame.spada:2015} studied this independently and found the same thing, and that this relationship could be crucial for modeling the slow rotator branch of period-mass diagrams. They discuss a trend suggesting that less massive stars experience stronger differential rotation (less core-envelope coupling). This would imply less efficient braking for lower mass stars as well. This is to say that the braking overall appears to be too strong in our current sub-solar mass models, and reducing core-envelope coupling in these models could make them overall match the slow rotator branch better. This effect would not help up simultaneously match the fast rotator branch; that would still be due to e.g., saturated stellar winds (under the \citealt{matt.etal:2015} model) or increased magnetic complexity (under the \citealt{garraffo.etal:2018} model). As this is not currently accounted for in our models (we assume a single $\nu_0$, i.e., core-envelope coupling level, for all stars), it could be another source of the discrepancy and should be accounted for in the future. Recently, \cite{spada.lanzafame:2020} have presented a model with mass scaling where core-envelope coupling decreases towards lower masses, and have been able to model the slow rotator branch very well. 
 A further physical explanation for this epoch of stalled spin down could be a period of greater magnetic field complexity, but ultimately any mechanism that dampens angular momentum loss rates could be at play, including, e.g., mass loss rates (as in \citealt{see.etal:2019}). However, if magnetic complexity were invoked, under the model of \cite{garraffo.etal:2018}, this would be reliant on the form of the function for $n$ (Eq. \ref{eq:g18.n}), which could have its mass dependence modified to allow this. Eq. \ref{eq:g18.n} was primarily formed to describe a general behavior of the models based on ZDI observations of stars, but is fairly unconstrained. If mass dependence were added, such that lower mass stars experience greater magnetic complexity after the age of the Praesepe, their spin down would be suppressed, achieving the observed behavior, similar to what is shown in our Fig. \ref{f:g18prot_biga}, for the case where $a$ has been increased from our default 0.03 to $a=0.05$. Recently \cite{see.etal:2019} have found that higher order modes in stellar magnetic fields may be more important at later ages as well, and \cite{vansaders.etal:2016, vansaders.pinsonneault.barbieri:2019} have found evidence for this too. If mass dependency in core-envelope coupling were introduced to our models, parameters would likely need to be re-calibrated, and it is unclear if a modified mass dependency in magnetic complexity would be required to reproduce the slow rotator branch; results from \cite{spada.lanzafame:2020} suggest that it may not be necessary for this purpose.

\subsection{Reproducing the solar lithium abundance}\label{ssec:disc.li7}

As seen in Figs. \ref{f:m15prot} and \ref{f:g18prot}, our solar mass models do not reproduce the solar lithium abundance, nor the observed dispersion in lithium abundances of solar-like stars \citep{carlos.etal:2019} under our default assumptions. This could be reconciled in at least two ways: either through prolonged rapid rotation through early ages, or strong differential rotation (as in \citealt{somers.pinsonneault:2016}). The historical arguments for rotationally induced mixing as a cause of lithium depletion involve several components. One is the observation that lithium depletion appears to follow a similar time dependence to stellar spin down \citep{skumanich:1972,charbonnel.vauclair.zahn:1992}. Additionally, the depletion has a dispersion that exists at fixed age, mass, and composition (as mentioned previously in Sec. \ref{ssec:res.li7}). In absence of rotational mixing, standard stellar models predict lithium depletion in PMS stars, with deep convective envelopes, and not on the main sequence for solar-like stars. However, lithium depletion has been observed in open clusters \citep{soderblom.etal:1993,sestito.randich:2005,thevenin.etal:2017,carlos.etal:2019}, in solar-like stars, taking place during the main sequence, calling for some mixing process (e.g., induced by rotation or other mechanisms) than standard stellar models operate with. Lithium is typically the most dramatically depleted light element in stellar surface abundances, but it is clear that some global mixing process is taking place within stars, as other light elements (boron and beryllium) can become depleted alongside lithium, especially in F- and B-class stars.

\begin{figure*}[ht!]
    \centering
    \includegraphics[width=\textwidth]{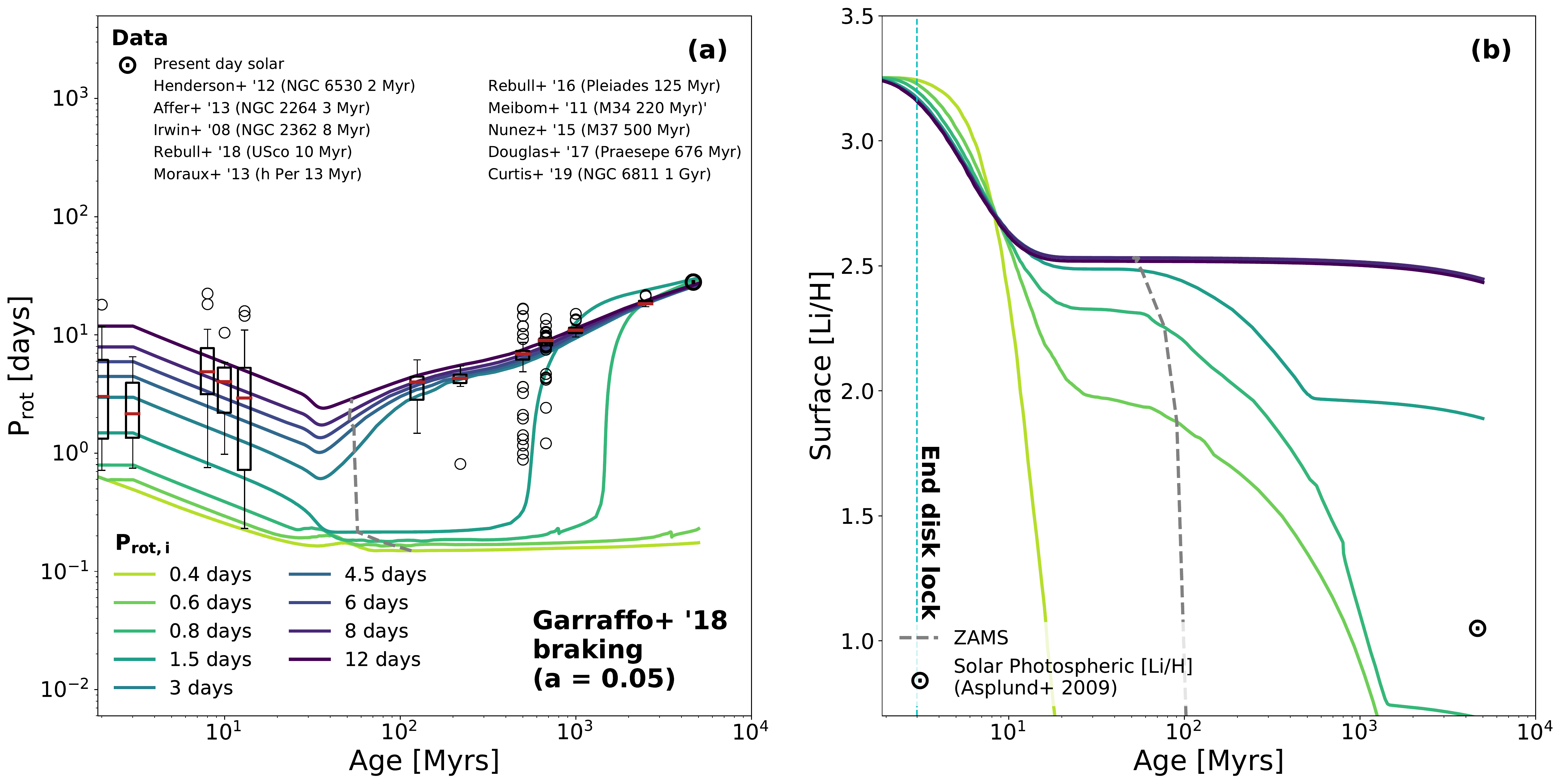}
    \caption{Similar to Figs. \ref{f:m15prot} and \ref{f:g18prot}, but just the evolution of (a) $\rm P_{rot}$ and (b) surface lithium abundance ($\rm log(^7Li/H) + 12$) for our solar model using the \cite{garraffo.etal:2018} braking model. We have changed $a$ from our default 0.03 to 0.05, delaying braking until later ages, and allowing greater rotational mixing for demonstrative purposes. The \cite{matt.etal:2015} could achieve this as well through a smaller $p$ value. Colors for each solid line are the same as described in Fig. \ref{f:m15prot}. Although effective in reproducing the solar lithium abundance, this scenario is not supported by observations. Note that the fastest rotators in panel (a) exhibit super-critical rotation, and so their behavior is erratic compared to our models with $P_{\rm{rot,i}}>1.5$ days.}
    \label{f:g18prot_biga}
\end{figure*}

To test whether rapid rotation could facilitate greater lithium depletion in our models, we have increased the constant $a$ in the \cite{garraffo.etal:2018} braking model (Eq. \ref{eq:g18.n}) to $a=0.05$ in Fig. \ref{f:g18prot_biga}, causing lower Rossby numbers to experience higher $n$ (and thus have their braking suppressed more strongly at early ages). In this case, our solar model with $P_{\rm{rot,i}}$ between 0.8 and 1.5 days appears to come close to reproducing the solar lithium abundance. In the case of \cite{garraffo.etal:2018} braking, it would then be higher order magnetic field contributions that suppress magnetic braking at early ages, leading to an extended period of rapid rotation. Our models under the \cite{matt.etal:2015} braking model could also achieve this through a lower $p$ value, weakening the braking model's dependence on $\rm R_o$, and keeping them spinning faster for longer. This scenario would suggest that the Sun began life as a rapid rotator. This scenario is uncertain with our models however, because these rapid rotators reach critical rotation ($\Omega/\Omega_c \approx 1$), where rotating \texttt{MESA} models become unreliable \citep{paxton.etal:2013,paxton.etal:2019}. More importantly, the scenario presented in Fig. \ref{f:g18prot_biga} does not seem to be supported by data.

However, for young stars of about solar mass and below, rapid rotators appear to be less depleted in lithium than slow rotators \citep{bouvier:2020}. Physical interpretations of this trend vary, with some citing inhibition to convective mixing with greater rotation rates \citep{baraffe.etal:2017}, or convective mixing inhibition from rotationally induced magnetic fields (e.g., \citealt{somers.pinsonneault:2014,somers.pinsonneault:2015,jeffries.etal:2017}), or perhaps star-disk interactions on the PMS \citep{bouvier:2008,eggenberger.etal:2012}. Whatever the case, observations say that rapid rotation early on leads to weaker lithium depletion, not stronger, indicating our models are incomplete in explaining this phenomenon correctly. 

Furthermore, as shown by \cite{delgado.etal:2014,carlos.etal:2019}, the majority of solar twins at the age of the Sun are more depleted in surface lithium more than our slowly rotating models predict. I.e., our slowly rotating models have surface lithium abundances that are too high. It then seems that observations of solar twins tell us that (regardless of $P_{\rm{rot,i}}$) all models need a boost in their mixing levels beyond what rotational mixing provides in order to match observations. These could come from various processes that alter the internal angular momentum transport (e.g., as explored by \citealt{somers.pinsonneault:2016}).

As found here and in \cite{amard.etal:2016,amard.etal:2019}, \cite{charbonnel.talon:2005,eggenberger.maeder.meynet:2005} noted that other mixing processes are needed to supplement rotationally enhanced mixing. Mixing mechanisms like gravity waves and magnetic fields do not directly mix material, but rather facilitate greater hydrodynamical transport in conjunction with rotation induced mixing. \cite{somers.pinsonneault:2016} parameterized these effects through the degree to which they enhance coupling between the stellar core and envelope via enhanced angular momentum transport (similar to the concept of our $\nu_0$ parameter). They found that solid body rotation (large $\nu_0$, and strong core-envelope coupling) paradoxically produces less efficient mixing than stars with differential rotation. They concluded that hybrid models that experience some degree of differential rotation (as observed in the Sun) are required to reproduce the solar lithium abundance. Too much differential rotation (small $\nu_0$) produces too weak of a core-envelope coupling for solar mass stars to converge to the tight sequence of slow rotation observed at late ages (e.g., the small spread in periods at late cluster ages \citealt{meibom.etal:2015,curtis.etal:2019}, and see the final columns of Figs. \ref{f:m15prot}, \ref{f:g18prot}). 

Our chosen default value of $\nu_0=2\times10^{4}\ \rm{cm^2}\ \rm{s^{-1}}$ creates differential rotation reminiscent of what is observed in the Sun, and produces a spread of rotation rates that appears to match the spread observed in open clusters over time in our solar mass model (Figs. \ref{f:m15prot} and \ref{f:g18prot}, panel (a)). If we were to decrease $\nu_0$ somewhat from our default value, we would induce stronger differential rotation, driving stronger shears, and more hydrodynamic transport (e.g., \citealt{somers.pinsonneault:2016}) between the core and envelope, producing stronger mixing. Processes that may alter the internal angular momentum transport (such as gravity waves or internal magnetic fields) provide another route for reproducing the solar lithium abundance that could work regardless of the adopted braking model. We have tried various values of $\nu_0$, but they had little effect in enhancing lithium depletion with our current models, unless we re-calibrate the constants $f_c$ and $f_\mu$ in the \cite{pinsonneault.etal:1989} diffusion approximation (see, \citealt{heger.langer.woosley:2000,choi.etal:2016} for additional details). In particular the parameter $f_c$, which scales the efficiency of composition mixing to angular momentum transport, could be adjusted (as in \citealt{somers.pinsonneault:2016}) to produce qualitatively and quantitatively different sets of model outcomes, with more or less lithium depletion. We save this re-calibration for future work.

While rotationally induced mixing may not be the direct driver of lithium depletion, rotation rate is still thought to be related to the depletion mechanism, and may account for the dispersion of abundances observed in solar twins, where the Sun appears to be especially low compared to most other stars \citep{carlos.etal:2019} at $\sim$4.6 Gyr. Observations by \cite{sestito.randich:2005} (also in \citealt{castro.etal:2016}) provide a map of how lithium abundance may evolve with time for various stars in open clusters, including near solar masses. A range of initial rotation rates amongst stars could explain this by varying the efficiency of the lithium depletion process(es). 

Assuming that clusters are born with some fraction of fast rotators (as in h Per and USco \citealt{moraux.etal:2013,somers.etal:2017,rebull.etal:2018}), and that these open clusters truly show us an evolutionary sequence of rotation rate with time (e.g., \citealt{fritzewski.etal:2020}, but c.f., \citealt{coker.pinsonneault.terndrup:2016}), observations suggest that clusters birth stars with a wide range of initial rotation periods. Observations e.g., from \cite{hartman.etal:2010,rebull.etal:2016} show that solar mass stars born with rapid rotation ($P_{\rm{rot,i}}\lesssim0.6$ days) are likely braked prior to $\sim$100 Myr, as very few such rapidly rotating stars near solar mass are observed beyond roughly this age in comparison to the distribution at 13 Myr \citep{moraux.etal:2013}. So, although a tight sequence of slow rotators is observed for solar masses at later ages, a wide range of initial rotation periods for solar mass stars could still have existed within a cluster at early ages, leaving an imprint on the subsequent dispersion of lithium abundances at late ages.

\subsection{The role of magnetic complexity}\label{ssec:disc.g18n}
In our modeling, under \cite{garraffo.etal:2018}, magnetic complexity serves the practical role of stalling angular momentum loss. There is evidence that the efficiency of angular momentum loss changes over time, as in \cite{agueros.etal:2018,curtis.etal:2019,see.etal:2019}, and possibly suggested by findings in \cite{vansaders.etal:2016, vansaders.pinsonneault.barbieri:2019}, but it is unclear why. Magnetic complexity provides a natural explanation (although core-envelope coupling timescales are another, as discussed by  \citealt{lanzafame.spada:2015,spada.lanzafame:2020}), but we do not fully understand how it should evolve throughout stellar evolution (i.e., in the present case whether Eq. \ref{eq:g18.n} properly mimics reality).

Presently, we have that $n$ evolves through the evolution of a star's $\rm R_o$, which tends to increase as stars get older. As stars spin down, $P_{\rm rot}$ increases, while along the main sequence, $\tau_c$ hardly varies. Higher mass stars have smaller convective envelopes, and smaller $\tau_c$ generally (see also the trend found in \citealt{wright.etal:2011}), meaning they arrive on the ZAMS with higher $\rm R_o$, tending to appear more as magnetic dipoles than low mass stars. Thus, higher mass stars tend to spin down at earlier ages than lower masses, as may be seen in Fig. \ref{f:m15prot}. This behavior helps our models using \cite{garraffo.etal:2018} braking achieve a collapsed sequence of slowly rotating stars more readily than when employing \cite{matt.etal:2015} braking, while also still predicting the presence of fast rotators with $\rm{M}<0.3\ \rm M_{\odot}$.

The behavior of $n$ at higher $\rm R_o$ according to Eq. \ref{eq:g18.n} is meant to produce a period of high magnetic complexity again under the calibration used in \cite{garraffo.etal:2018} (at what would tend to be later ages). The stalled angular momentum loss brought about by this could be an additional strength of including magnetic complexity in magnetic braking models. In the present case, we may need a modified mass dependence on the term $a$ in Eq. \ref{eq:g18.n}, such that lower mass stars slow down, but re-enter a period of high magnetic complexity (suppressed angular momentum loss) again at lower $\rm R_o$ than higher mass stars do. In other words, the function in Fig. \ref{f:g18n} may need to shift leftwards, towards smaller $\rm R_o$ as mass decreases. It is also worth noting that at low enough Rossby numbers the saturated regime (described in Sec. \ref{ssec:m15form}) may transition to a ``super-saturated'' regime \citep{james.etal:2000,wright.etal:2011,jeffries.etal:2011,argiroffi.etal:2016}. Evidence for super-saturation is limited to a fairly low number of stars, but if the rapid rotation of super-saturated stars is due to inefficient angular momentum loss, the implicit link would be that super-saturation may be related to higher magnetic field complexity under the \cite{garraffo.etal:2018} model, which is partially based on ZDI data that says magnetic field complexity increases with increasing rotation rate in active stars. Other dependencies of the angular momentum loss rate and magnetic field strength on stellar properties, such as atmospheric pressure and mass have been explored (e.g., by \citealt{gallet.bouvier:2013} and \citealt{vansaders.pinsonneault:2013}), and could provide additional physical updates to these models in the future. Whether higher order modes should play a significant role in modeling angular momentum loss is currently uncertain, as \cite{see.etal:2019,see.etal:2020} have shown that a majority of stars may not be significantly affected by high order magnetic field topologies, except perhaps at late ages.

The results of \cite{garraffo.etal:2018} were based on single mode MHD simulations, to conceptualize the effect of high order magnetic field topology on outgoing stellar winds \citep{reville.etal:2015,garraffo.drake.cohen:2015,garraffo.drake.cohen:2016}. \cite{finley.matt:2017,finley.matt:2018} conducted MHD simulations with mixed magnetic field modes (as would be found in nature) to study the extent to which higher order magnetic fields may affect outgoing stellar winds under more realistic conditions. They found that generally, the dipolar mode dominates the wind morphology for a wide range of mixed magnetic field topology configurations, but that certain configurations do show that high order magnetic fields dominate the wind morphology, so long as magnetic energy is concentrated moreso in the higher orders. 

\cite{garraffo.drake.cohen:2016}  compared their approximations to ZDI maps and MHD simulations for several real stars. They found very good agreement between the mass and angular momentum loss rates for the observed ZDI map MHD simulations and those of the pure modes that corresponded to the dominant magnetic field order of the observed ZDI maps. The decomposed observed ZDI maps showed fields dominated by higher order magnetic fields ($n>1$, i.e., non-dipolar). 

As analyzed by \cite{lehman.etal:2019}, it appears to be difficult to accurately recover the magnetic fluxes that may be associated with the dipolar, quadrupolar, and octopolar magnetic field orders via ZDI due to the limits of spectral resolution. Given these difficulties, it is not easy to understand the partition of total magnetic flux among different field orders, and the true influence of higher order magnetic fields on actual stellar wind and angular momentum loss rates.

If, as suggested by \cite{see.etal:2019,see.etal:2020}, magnetic flux is primarily distributed in the dipole mode for a majority of stars, then high order magnetic complexity may play a reduced role on angular momentum loss rates. \cite{see.etal:2020} found that below a certain threshold of mass loss, non-dipolar magnetic fields negligibly influence angular momentum loss. They estimate mass loss rates for a sample of stars, and find many to be below this threshold, but note that estimating the mass loss rates is subject to significant uncertainty. Furthermore, \cite{garraffo.drake.cohen:2015,garraffo.drake.cohen:2016,finley.matt:2017,garraffo.etal:2018,finley.matt:2018} point out that high order magnetic fields can reduce mass loss rates themselves (if they dominate the magnetic flux). Thus, it is not clear if estimated mass loss rates might be low because of high order magnetic fields in the first place. If high order magnetic fields comprise a dominant fraction of the magnetic field energy compared to the dipole mode, high order magnetic fields should have some effect on the angular momentum loss rate, and magnetic complexity should not be ignored in modeling magnetic braking.

\subsection{Super-critical rotation}\label{ssec:disc.fastp}
As mentioned in Sec. \ref{ssec:ini}, we have calculated faster rotation rates ($P_{\rm{rot,i}}= 0.4, 0.6$ and 0.8 days) than what is shown in the bulk of our results. We have excluded these rapidly rotating models for the most part because near solar masses, they tend to reach super-critical rotation rates, at which point calculations of rotating models in \texttt{MESA} become uncertain. This does not necessarily mean that these stars would not exist in nature; indeed some solar mass stars rotating this fast are observed in the data of \cite{irwin.etal:2008,henderson.etal:2012,affer.etal:2013}, in clusters 2-8 Myr old or so. The number of rapidly rotating  solar mass stars appears to diminish between about 13 and 100 Myr \citep{moraux.etal:2013}. It would be straightforward to assume that the reason is because of magnetic braking, in lieu of a confident modeling of stellar evolution under critical rotation.

There is evidence that solar mass rapid rotators cease to rotate rapidly in open cluster rotation period data between $\sim$13 and $\sim$100 Myr(e.g., \citealt{hartman.etal:2010,meibom.etal:2011,meibom.etal:2015,nunez.etal:2015,rebull.etal:2016,douglas.etal:2017,curtis.etal:2019}, c.f., at early ages \citealt{moraux.etal:2013}). It is likely that the rapid rotators have simply been braked, as rapidly rotating stars may be necessary to explain the observed dispersion in stellar lithium abundances \citep{sestito.randich:2005,delgado.etal:2014,castro.etal:2016,carlos.etal:2019}. As discussed above in Sect.~\ref{ssec:disc.li7}, rotation is thought to be linked to the efficiency of the Li depletion mechanism \citep{bouvier:2020}. Likely then, the braking of these rapid rotators takes place somewhere between 13 and 100 Myr; presumably, the more slowly rotating stars begin braking in a similar time frame. Our default implementations of \cite{matt.etal:2015} and \cite{garraffo.etal:2018} braking achieve this, rather than e.g., the scenario presented in our Fig. \ref{f:g18prot_biga}, though our fastest rotators remain unstable. We could tune the parameters in our implementations of the braking models to prevent our rapidly rotating solar mass stars from reaching critical rotation, but opt not to because they are not physically restricted from doing so, and they are not highly impactful on our results.


\section{Conclusions} \label{sec:conclusions}

We have implemented the magnetic braking models of \cite{matt.etal:2015} and \cite{garraffo.etal:2018} in the MIST framework of the MESA stellar evolution code. Our goal was to test their predictions against observed stellar rotation period data and examine some of their implications for some aspects of stellar physics associated with rotation.

In terms of practicality, we find that the \cite{matt.etal:2015} braking model provides a good overall description of stellar spin down with our models. However, we also see potential for the inclusion of magnetic complexity, as in the \cite{garraffo.etal:2018} braking model to improve our modeling abilities. The \cite{garraffo.etal:2018} formalism captures the bimodality of the observed rotation period distributions quite well, while the \cite{matt.etal:2015} model does not. At old ages, both braking models predict rotation that is too slow, while prior to this, the \cite{matt.etal:2015} model tends to predict rotation that is too fast for masses $0.3\ \rm M_{\odot}\lesssim M\lesssim 1\ \rm M_{\odot}$, and the \cite{garraffo.etal:2018} formalism predict rotation too slow in that mass regime at 600 Myr onwards. In spite of these shortcomings, it seems that magnetic complexity could play a unique role in characterizing the observed bimodal rotation period distributions in clusters younger than about $1$ Gyr.

 That both braking models tend to predict rotation that is too slow at ages $> 1$ Gyr, seems to suggest a missing mass dependant effect in the evolution of $P_{\rm rot}$in our models. Aside from explaining the observed bimodal distributions of low mass stars, increased magnetic complexity is a mechanism through which angular momentum loss may be halted by suppressing magnetic braking at late ages, keeping stars from spinning down too much. Recent findings suggest that a halt to angular momentum loss may be necessary at these late ages \citep{vansaders.etal:2016,agueros.etal:2018,curtis.etal:2019,vansaders.pinsonneault.barbieri:2019}. At least two mechanisms that could provide a mass dependent effect are a modified dependency in magnetic braking, or a mass dependency in core-envelope coupling time scales \citep{lanzafame.spada:2015,somers.pinsonneault:2016,spada.lanzafame:2020}. Of course, both effects could act on stars, and their interplay should be studied and constrained in future work on resolving spin down at ages greater than 1 Gyr. Thus, we plan to incorporate mass dependence in core-envelope re-coupling in future models to study these effects.

Studying the solar lithium abundance has provided a useful constraint on mixing efficiencies in these models, and in constraining the spin down rate of our models. Our default models show lithium depletion, but it is not sufficient to reproduce the solar lithium abundance. Extended periods of rapid rotation at early ages allow for a sufficient amount of lithium depletion in our solar model via enhanced rotational mixing. However, observations make this scenario appear unlikely. In regards to solar lithium depletion, there then appears to be a mixing process common to all solar mass stars (slow and rapid rotation) that is missing. E.g., \cite{carlos.etal:2019} show that solar mass stars are generally more depleted than the majority of our models predict. Models may benefit from additional sources of mixing, such as gravity waves, or internal magnetic fields \citep{charbonnel.talon:2005,eggenberger.maeder.meynet:2005,denissenkov.etal:2010,somers.pinsonneault:2016}, and/or other effects (e.g., variable convective mixing efficiencies \citealt{baraffe.etal:2017} linked to rotation rate).

We have laid groundwork for the extension of \texttt{MIST} \citep{dotter:2016,choi.etal:2016,gossage.etal:2018} model grids here with implementations of magnetic braking, important for allowing our rotating model grids to cover the complete stellar mass range. In future work, we plan to release an extended grid of models within the \texttt{MIST} framework utilizing the rotation methodology presented here.

\acknowledgments

S.G. acknowledges the National Science Foundation Graduate Research Fellowship under grant No. DGE1745303. A.D. is supported by the National Aeronautics and Space Administration (NASA) under contract No. NNG16PJ26C issued through the WFIRST Science Investigation Teams Program. J.J.D. was funded by NASA contract NAS8-03060 to the CXC and thanks the Director, Pat Slane, for advice and support. The authors thank Yaroslav Lazovik of Moscow State University for contributing corrections to this paper. We would also like to thank Bill Paxton and the \texttt{MESA} community for making this work possible.

%

//
\software{\texttt{MESA r11701} \citep{paxton.etal:2011,paxton.etal:2013,paxton.etal:2015, paxton.etal:2018, paxton.etal:2019}}





\begin{thebibliography}{}
\expandafter\ifx\csname natexlab\endcsname\relax\def\natexlab#1{#1}\fi
\providecommand{\url}[1]{\href{#1}{#1}}
\providecommand{\dodoi}[1]{doi:~\href{http://doi.org/#1}{\nolinkurl{#1}}}
\providecommand{\doeprint}[1]{\href{http://ascl.net/#1}{\nolinkurl{http://ascl.net/#1}}}
\providecommand{\doarXiv}[1]{\href{https://arxiv.org/abs/#1}{\nolinkurl{https://arxiv.org/abs/#1}}}

\bibitem[{{Affer} {et~al.}(2013){Affer}, {Micela}, {Favata}, {Flaccomio}, \&
  {Bouvier}}]{affer.etal:2013}
{Affer}, L., {Micela}, G., {Favata}, F., {Flaccomio}, E., \& {Bouvier}, J.
  2013, \mnras, 430, 1433, \dodoi{10.1093/mnras/stt003}

\bibitem[{{Ag{\"u}eros} {et~al.}(2018){Ag{\"u}eros}, {Bowsher}, {Bochanski},
  {Cargile}, {Covey}, {Douglas}, {Kraus}, {Kundert}, {Law}, {Ahmadi}, \&
  {Arce}}]{agueros.etal:2018}
{Ag{\"u}eros}, M.~A., {Bowsher}, E.~C., {Bochanski}, J.~J., {et~al.} 2018,
  \apj, 862, 33, \dodoi{10.3847/1538-4357/aac6ed}

\bibitem[{{Alvarado-G{\'o}mez} {et~al.}(2015){Alvarado-G{\'o}mez}, {Hussain},
  {Grunhut}, {Fares}, {Donati}, {Alecian}, {Kochukhov}, {Oksala}, {Morin},
  {Redfield}, {Cohen}, {Drake}, {Jardine}, {Matt}, {Petit}, \&
  {Walter}}]{alvarado-gomez.etal:2015}
{Alvarado-G{\'o}mez}, J.~D., {Hussain}, G.~A.~J., {Grunhut}, J., {et~al.} 2015,
  \aap, 582, A38, \dodoi{10.1051/0004-6361/201525771}

\bibitem[{{Amard} {et~al.}(2016){Amard}, {Palacios}, {Charbonnel}, {Gallet}, \&
  {Bouvier}}]{amard.etal:2016}
{Amard}, L., {Palacios}, A., {Charbonnel}, C., {Gallet}, F., \& {Bouvier}, J.
  2016, \aap, 587, A105, \dodoi{10.1051/0004-6361/201527349}

\bibitem[{{Amard} {et~al.}(2019){Amard}, {Palacios}, {Charbonnel}, {Gallet},
  {Georgy}, {Lagarde}, \& {Siess}}]{amard.etal:2019}
{Amard}, L., {Palacios}, A., {Charbonnel}, C., {et~al.} 2019, \aap, 631, A77,
  \dodoi{10.1051/0004-6361/201935160}

\bibitem[{{Angus} {et~al.}(2015){Angus}, {Aigrain}, {Foreman-Mackey}, \&
  {McQuillan}}]{angus.etal:2015}
{Angus}, R., {Aigrain}, S., {Foreman-Mackey}, D., \& {McQuillan}, A. 2015,
  \mnras, 450, 1787, \dodoi{10.1093/mnras/stv423}

\bibitem[{{Argiroffi} {et~al.}(2016){Argiroffi}, {Caramazza}, {Micela},
  {Sciortino}, {Moraux}, {Bouvier}, \& {Flaccomio}}]{argiroffi.etal:2016}
{Argiroffi}, C., {Caramazza}, M., {Micela}, G., {et~al.} 2016, \aap, 589, A113,
  \dodoi{10.1051/0004-6361/201526539}

\bibitem[{{Asplund} {et~al.}(2009){Asplund}, {Grevesse}, {Sauval}, \&
  {Scott}}]{asplund.etal:2009}
{Asplund}, M., {Grevesse}, N., {Sauval}, A.~J., \& {Scott}, P. 2009, \araa, 47,
  481, \dodoi{10.1146/annurev.astro.46.060407.145222}

\bibitem[{{Baraffe} {et~al.}(1998){Baraffe}, {Chabrier}, {Allard}, \&
  {Hauschildt}}]{baraffe.etal:1998}
{Baraffe}, I., {Chabrier}, G., {Allard}, F., \& {Hauschildt}, P.~H. 1998, \aap,
  337, 403.
\newblock \doarXiv{astro-ph/9805009}

\bibitem[{{Baraffe} {et~al.}(2017){Baraffe}, {Pratt}, {Goffrey}, {Constantino},
  {Folini}, {Popov}, {Walder}, \& {Viallet}}]{baraffe.etal:2017}
{Baraffe}, I., {Pratt}, J., {Goffrey}, T., {et~al.} 2017, \apjl, 845, L6,
  \dodoi{10.3847/2041-8213/aa82ff}

\bibitem[{{Barnes}(2003)}]{barnes:2003}
{Barnes}, S.~A. 2003, \apj, 586, 464, \dodoi{10.1086/367639}

\bibitem[{{Beck} {et~al.}(2017){Beck}, {do Nascimento}, {Duarte}, {Salabert},
  {Tkachenko}, {Mathis}, {Mathur}, {Garc{\'\i}a}, {Castro}, {Pall{\'e}},
  {Egeland }, {Montes}, {Creevey}, {Andersen}, {Kamath}, \& {van
  Winckel}}]{beck.etal:2017}
{Beck}, P.~G., {do Nascimento}, J.~D., J., {Duarte}, T., {et~al.} 2017, \aap,
  602, A63, \dodoi{10.1051/0004-6361/201629820}

\bibitem[{{B{\"o}hm-Vitense}(1958)}]{bohm-vitense:1958}
{B{\"o}hm-Vitense}, E. 1958, \zap, 46, 108

\bibitem[{{Bouvier}(2008)}]{bouvier:2008}
{Bouvier}, J. 2008, \aap, 489, L53, \dodoi{10.1051/0004-6361:200810574}

\bibitem[{{Bouvier}(2020)}]{bouvier:2020}
---. 2020, \memsai, 91, 39.
\newblock \doarXiv{2009.02086}

\bibitem[{{Bouvier} {et~al.}(2014){Bouvier}, {Matt}, {Mohanty}, {Scholz},
  {Stassun}, \& {Zanni}}]{bouvier.etal:2014}
{Bouvier}, J., {Matt}, S.~P., {Mohanty}, S., {et~al.} 2014, in Protostars and
  Planets VI, ed. H.~{Beuther}, R.~S. {Klessen}, C.~P. {Dullemond}, \&
  T.~{Henning}, 433, \dodoi{10.2458/azu_uapress_9780816531240-ch019}

\bibitem[{{Brandt}(1966)}]{brandt:1966}
{Brandt}, J.~C. 1966, \apj, 144, 1221, \dodoi{10.1086/148720}

\bibitem[{{Brandt} \& {Huang}(2015)}]{brandt.huang:2015a}
{Brandt}, T.~D., \& {Huang}, C.~X. 2015, \apj, 807, 58,
  \dodoi{10.1088/0004-637X/807/1/58}

\bibitem[{{Bressan} {et~al.}(2012){Bressan}, {Marigo}, {Girardi}, {Salasnich},
  {Dal Cero}, {Rubele}, \& {Nanni}}]{bressan.etal:2012}
{Bressan}, A., {Marigo}, P., {Girardi}, L., {et~al.} 2012, \mnras, 427, 127,
  \dodoi{10.1111/j.1365-2966.2012.21948.x}

\bibitem[{{Carlos} {et~al.}(2019){Carlos}, {Mel{\'e}ndez}, {Spina}, {dos
  Santos}, {Bedell}, {Ramirez}, {Asplund}, {Bean}, {Yong}, {Yana Galarza}, \&
  {Alves-Brito}}]{carlos.etal:2019}
{Carlos}, M., {Mel{\'e}ndez}, J., {Spina}, L., {et~al.} 2019, \mnras, 485,
  4052, \dodoi{10.1093/mnras/stz681}

\bibitem[{{Castro} {et~al.}(2016){Castro}, {Duarte}, {Pace}, \& {do
  Nascimento}}]{castro.etal:2016}
{Castro}, M., {Duarte}, T., {Pace}, G., \& {do Nascimento}, J.~D. 2016, \aap,
  590, A94, \dodoi{10.1051/0004-6361/201527583}

\bibitem[{{Charbonnel} \& {Talon}(2005)}]{charbonnel.talon:2005}
{Charbonnel}, C., \& {Talon}, S. 2005, Science, 309, 2189,
  \dodoi{10.1126/science.1116849}

\bibitem[{{Charbonnel} {et~al.}(1994){Charbonnel}, {Vauclair}, {Maeder},
  {Meynet}, \& {Schaller}}]{charbonnel.etal:1994}
{Charbonnel}, C., {Vauclair}, S., {Maeder}, A., {Meynet}, G., \& {Schaller}, G.
  1994, \aap, 283, 155

\bibitem[{{Charbonnel} {et~al.}(1992){Charbonnel}, {Vauclair}, \&
  {Zahn}}]{charbonnel.vauclair.zahn:1992}
{Charbonnel}, C., {Vauclair}, S., \& {Zahn}, J.~P. 1992, \aap, 255, 191

\bibitem[{{Charbonnel} {et~al.}(2017){Charbonnel}, {Decressin}, {Lagarde},
  {Gallet}, {Palacios}, {Auri{\`e}re}, {Konstantinova-Antova}, {Mathis},
  {Anderson}, \& {Dintrans}}]{charbonnel.etal:2017}
{Charbonnel}, C., {Decressin}, T., {Lagarde}, N., {et~al.} 2017, \aap, 605,
  A102, \dodoi{10.1051/0004-6361/201526724}

\bibitem[{{Choi} {et~al.}(2016){Choi}, {Dotter}, {Conroy}, {Cantiello},
  {Paxton}, \& {Johnson}}]{choi.etal:2016}
{Choi}, J., {Dotter}, A., {Conroy}, C., {et~al.} 2016, \apj, 823, 102,
  \dodoi{10.3847/0004-637X/823/2/102}

\bibitem[{{Coker} {et~al.}(2016){Coker}, {Pinsonneault}, \&
  {Terndrup}}]{coker.pinsonneault.terndrup:2016}
{Coker}, C.~T., {Pinsonneault}, M., \& {Terndrup}, D.~M. 2016, \apj, 833, 122,
  \dodoi{10.3847/1538-4357/833/1/122}

\bibitem[{{Couvidat} {et~al.}(2003){Couvidat}, {Garc{\'\i}a},
  {Turck-Chi{\`e}ze}, {Corbard}, {Henney}, \&
  {Jim{\'e}nez-Reyes}}]{couvidat.etal:2003}
{Couvidat}, S., {Garc{\'\i}a}, R.~A., {Turck-Chi{\`e}ze}, S., {et~al.} 2003,
  \apjl, 597, L77, \dodoi{10.1086/379698}

\bibitem[{{Cranmer} \& {Saar}(2011)}]{cranmer.saar:2011}
{Cranmer}, S.~R., \& {Saar}, S.~H. 2011, \apj, 741, 54,
  \dodoi{10.1088/0004-637X/741/1/54}

\bibitem[{{Currie} {et~al.}(2007){Currie}, {Kenyon}, {Balog}, {Bragg}, \&
  {Tokarz}}]{currie.etal:2007}
{Currie}, T., {Kenyon}, S.~J., {Balog}, Z., {Bragg}, A., \& {Tokarz}, S. 2007,
  \apjl, 669, L33, \dodoi{10.1086/523595}

\bibitem[{{Curtis} {et~al.}(2019){Curtis}, {Ag{\"u}eros}, {Douglas}, \&
  {Meibom}}]{curtis.etal:2019}
{Curtis}, J.~L., {Ag{\"u}eros}, M.~A., {Douglas}, S.~T., \& {Meibom}, S. 2019,
  \apj, 879, 49, \dodoi{10.3847/1538-4357/ab2393}

\bibitem[{{Delgado Mena} {et~al.}(2014){Delgado Mena}, {Israelian},
  {Gonz{\'a}lez Hern{\'a}ndez}, {Sousa}, {Mortier}, {Santos}, {Adibekyan},
  {Fernand es}, {Rebolo}, {Udry}, \& {Mayor}}]{delgado.etal:2014}
{Delgado Mena}, E., {Israelian}, G., {Gonz{\'a}lez Hern{\'a}ndez}, J.~I.,
  {et~al.} 2014, \aap, 562, A92, \dodoi{10.1051/0004-6361/201321493}

\bibitem[{{Demarque} {et~al.}(2004){Demarque}, {Woo}, {Kim}, \&
  {Yi}}]{demarque.etal:2004}
{Demarque}, P., {Woo}, J.-H., {Kim}, Y.-C., \& {Yi}, S.~K. 2004, \apjs, 155,
  667, \dodoi{10.1086/424966}

\bibitem[{{Denissenkov} \&
  {Pinsonneault}(2007)}]{denissenkov.pinsonneault:2007}
{Denissenkov}, P.~A., \& {Pinsonneault}, M. 2007, \apj, 655, 1157,
  \dodoi{10.1086/510345}

\bibitem[{{Denissenkov} {et~al.}(2010){Denissenkov}, {Pinsonneault},
  {Terndrup}, \& {Newsham}}]{denissenkov.etal:2010}
{Denissenkov}, P.~A., {Pinsonneault}, M., {Terndrup}, D.~M., \& {Newsham}, G.
  2010, \apj, 716, 1269, \dodoi{10.1088/0004-637X/716/2/1269}

\bibitem[{{Donati} \& {Landstreet}(2009)}]{donati.landstreet:2009}
{Donati}, J.~F., \& {Landstreet}, J.~D. 2009, \araa, 47, 333,
  \dodoi{10.1146/annurev-astro-082708-101833}

\bibitem[{{Dotter}(2016)}]{dotter:2016}
{Dotter}, A. 2016, \apjs, 222, 8, \dodoi{10.3847/0067-0049/222/1/8}

\bibitem[{{Dotter} {et~al.}(2008){Dotter}, {Chaboyer}, {Jevremovi{\'c}},
  {Kostov}, {Baron}, \& {Ferguson}}]{dotter.etal:2008}
{Dotter}, A., {Chaboyer}, B., {Jevremovi{\'c}}, D., {et~al.} 2008, \apjs, 178,
  89, \dodoi{10.1086/589654}

\bibitem[{{Douglas} {et~al.}(2017){Douglas}, {Ag{\"u}eros}, {Covey}, \&
  {Kraus}}]{douglas.etal:2017}
{Douglas}, S.~T., {Ag{\"u}eros}, M.~A., {Covey}, K.~R., \& {Kraus}, A. 2017,
  \apj, 842, 83, \dodoi{10.3847/1538-4357/aa6e52}

\bibitem[{{Eggenberger} {et~al.}(2012){Eggenberger}, {Haemmerl{\'e}}, {Meynet},
  \& {Maeder}}]{eggenberger.etal:2012}
{Eggenberger}, P., {Haemmerl{\'e}}, L., {Meynet}, G., \& {Maeder}, A. 2012,
  \aap, 539, A70, \dodoi{10.1051/0004-6361/201118432}

\bibitem[{{Eggenberger} {et~al.}(2005){Eggenberger}, {Maeder}, \&
  {Meynet}}]{eggenberger.maeder.meynet:2005}
{Eggenberger}, P., {Maeder}, A., \& {Meynet}, G. 2005, \aap, 440, L9,
  \dodoi{10.1051/0004-6361:200500156}

\bibitem[{{Eggenberger} {et~al.}(2010){Eggenberger}, {Maeder}, \&
  {Meynet}}]{eggenberger.maeder.meynet:2010}
---. 2010, \aap, 519, L2, \dodoi{10.1051/0004-6361/201014939}

\bibitem[{{Eggenberger} {et~al.}(2008){Eggenberger}, {Meynet}, {Maeder},
  {Hirschi}, {Charbonnel}, {Talon}, \& {Ekstr{\"o}m}}]{eggenberger:2008}
{Eggenberger}, P., {Meynet}, G., {Maeder}, A., {et~al.} 2008, \apss, 316, 43,
  \dodoi{10.1007/s10509-007-9511-y}

\bibitem[{{Ekstr{\"o}m} {et~al.}(2012){Ekstr{\"o}m}, {Georgy}, {Eggenberger},
  {Meynet}, {Mowlavi}, {Wyttenbach}, {Granada}, {Decressin}, {Hirschi},
  {Frischknecht}, {Charbonnel}, \& {Maeder}}]{ekstrom.etal:2012}
{Ekstr{\"o}m}, S., {Georgy}, C., {Eggenberger}, P., {et~al.} 2012, \aap, 537,
  A146, \dodoi{10.1051/0004-6361/201117751}

\bibitem[{{Eldridge} {et~al.}(2017){Eldridge}, {Stanway}, {Xiao}, {McClelland},
  {Taylor}, {Ng}, {Greis}, \& {Bray}}]{eldrige.etal:2017}
{Eldridge}, J.~J., {Stanway}, E.~R., {Xiao}, L., {et~al.} 2017, \pasa, 34,
  e058, \dodoi{10.1017/pasa.2017.51}

\bibitem[{{Endal} \& {Sofia}(1976)}]{endal.sofia:1976}
{Endal}, A.~S., \& {Sofia}, S. 1976, \apj, 210, 184, \dodoi{10.1086/154817}

\bibitem[{{Endal} \& {Sofia}(1978)}]{endal.sofia:1978}
---. 1978, \apj, 220, 279, \dodoi{10.1086/155904}

\bibitem[{{Espinosa Lara} \& {Rieutord}(2011)}]{espinosalara.rieutord:2011}
{Espinosa Lara}, F., \& {Rieutord}, M. 2011, \aap, 533, A43,
  \dodoi{10.1051/0004-6361/201117252}

\bibitem[{{Fedele} {et~al.}(2010){Fedele}, {van den Ancker}, {Henning},
  {Jayawardhana}, \& {Oliveira}}]{fedele.etal:2010}
{Fedele}, D., {van den Ancker}, M.~E., {Henning}, T., {Jayawardhana}, R., \&
  {Oliveira}, J.~M. 2010, \aap, 510, A72, \dodoi{10.1051/0004-6361/200912810}

\bibitem[{{Finley} \& {Matt}(2017)}]{finley.matt:2017}
{Finley}, A.~J., \& {Matt}, S.~P. 2017, \apj, 845, 46,
  \dodoi{10.3847/1538-4357/aa7fb9}

\bibitem[{{Finley} \& {Matt}(2018)}]{finley.matt:2018}
---. 2018, \apj, 854, 78, \dodoi{10.3847/1538-4357/aaaab5}

\bibitem[{{Fritzewski} {et~al.}(2020){Fritzewski}, {Barnes}, {James}, \&
  {Strassmeier}}]{fritzewski.etal:2020}
{Fritzewski}, D.~J., {Barnes}, S.~A., {James}, D.~J., \& {Strassmeier}, K.~G.
  2020, \aap, 641, A51, \dodoi{10.1051/0004-6361/201936860}

\bibitem[{{Gallet} \& {Bouvier}(2013)}]{gallet.bouvier:2013}
{Gallet}, F., \& {Bouvier}, J. 2013, \aap, 556, A36,
  \dodoi{10.1051/0004-6361/201321302}

\bibitem[{{Gallet} \& {Bouvier}(2015)}]{gallet.bouvier:2015}
---. 2015, \aap, 577, A98, \dodoi{10.1051/0004-6361/201525660}

\bibitem[{{Gallet} {et~al.}(2019){Gallet}, {Zanni}, \&
  {Amard}}]{gallet.zanni.amard:2019}
{Gallet}, F., {Zanni}, C., \& {Amard}, L. 2019, \aap, 632, A6,
  \dodoi{10.1051/0004-6361/201935432}

\bibitem[{{Garraffo} {et~al.}(2015){Garraffo}, {Drake}, \&
  {Cohen}}]{garraffo.drake.cohen:2015}
{Garraffo}, C., {Drake}, J.~J., \& {Cohen}, O. 2015, \apj, 813, 40,
  \dodoi{10.1088/0004-637X/813/1/40}

\bibitem[{{Garraffo} {et~al.}(2016){Garraffo}, {Drake}, \&
  {Cohen}}]{garraffo.drake.cohen:2016}
---. 2016, \aap, 595, A110, \dodoi{10.1051/0004-6361/201628367}

\bibitem[{{Garraffo} {et~al.}(2018){Garraffo}, {Drake}, {Dotter}, {Choi},
  {Burke}, {Moschou}, {Alvarado-G{\'o}mez}, {Kashyap}, \&
  {Cohen}}]{garraffo.etal:2018}
{Garraffo}, C., {Drake}, J.~J., {Dotter}, A., {et~al.} 2018, \apj, 862, 90,
  \dodoi{10.3847/1538-4357/aace5d}

\bibitem[{{Georgy} {et~al.}(2014){Georgy}, {Granada}, {Ekstr{\"o}m}, {Meynet},
  {Anderson}, {Wyttenbach}, {Eggenberger}, \& {Maeder}}]{georgy.etal:2014}
{Georgy}, C., {Granada}, A., {Ekstr{\"o}m}, S., {et~al.} 2014, \aap, 566, A21,
  \dodoi{10.1051/0004-6361/201423881}

\bibitem[{{Georgy} {et~al.}(2019){Georgy}, {Charbonnel}, {Amard}, {Bastian},
  {Ekstr{\"o}m}, {Lardo}, {Palacios}, {Eggenberger}, {Cabrera-Ziri}, {Gallet},
  \& {Lagarde}}]{georgy.etal:2019}
{Georgy}, C., {Charbonnel}, C., {Amard}, L., {et~al.} 2019, \aap, 622, A66,
  \dodoi{10.1051/0004-6361/201834505}

\bibitem[{{Gilliland}(1985)}]{gilliland:1985}
{Gilliland}, R.~L. 1985, \apj, 299, 286, \dodoi{10.1086/163699}

\bibitem[{{Gossage} {et~al.}(2018){Gossage}, {Conroy}, {Dotter}, {Choi},
  {Rosenfield}, {Cargile}, \& {Dolphin}}]{gossage.etal:2018}
{Gossage}, S., {Conroy}, C., {Dotter}, A., {et~al.} 2018, \apj, 863, 67,
  \dodoi{10.3847/1538-4357/aad0a0}

\bibitem[{{Gossage} {et~al.}(2019){Gossage}, {Conroy}, {Dotter},
  {Cabrera-Ziri}, {Dolphin}, {Bastian}, {Dalcanton}, {Goudfrooij}, {Johnson},
  {Williams}, {Rosenfield}, {Kalirai}, \& {Fouesneau}}]{gossage.etal:2019}
---. 2019, \apj, 887, 199, \dodoi{10.3847/1538-4357/ab5717}

\bibitem[{{Hartman} {et~al.}(2010){Hartman}, {Bakos}, {Kov{\'a}cs}, \&
  {Noyes}}]{hartman.etal:2010}
{Hartman}, J.~D., {Bakos}, G.~{\'A}., {Kov{\'a}cs}, G., \& {Noyes}, R.~W. 2010,
  \mnras, 408, 475, \dodoi{10.1111/j.1365-2966.2010.17147.x}

\bibitem[{{Heger} \& {Langer}(2000)}]{heger.langer:2000}
{Heger}, A., \& {Langer}, N. 2000, \apj, 544, 1016, \dodoi{10.1086/317239}

\bibitem[{{Heger} {et~al.}(2000){Heger}, {Langer}, \&
  {Woosley}}]{heger.langer.woosley:2000}
{Heger}, A., {Langer}, N., \& {Woosley}, S.~E. 2000, \apj, 528, 368,
  \dodoi{10.1086/308158}

\bibitem[{{Henderson} \& {Stassun}(2012)}]{henderson.etal:2012}
{Henderson}, C.~B., \& {Stassun}, K.~G. 2012, \apj, 747, 51,
  \dodoi{10.1088/0004-637X/747/1/51}

\bibitem[{{Howe}(2009)}]{howe:2009}
{Howe}, R. 2009, Living Reviews in Solar Physics, 6, 1,
  \dodoi{10.12942/lrsp-2009-1}

\bibitem[{{Irwin} {et~al.}(2011){Irwin}, {Berta}, {Burke}, {Charbonneau},
  {Nutzman}, {West}, \& {Falco}}]{irwin.etal:2011}
{Irwin}, J., {Berta}, Z.~K., {Burke}, C.~J., {et~al.} 2011, \apj, 727, 56,
  \dodoi{10.1088/0004-637X/727/1/56}

\bibitem[{{Irwin} \& {Bouvier}(2009)}]{irwin.bouvier:2009}
{Irwin}, J., \& {Bouvier}, J. 2009, in The Ages of Stars, ed. E.~E. {Mamajek},
  D.~R. {Soderblom}, \& R.~F.~G. {Wyse}, Vol. 258, 363--374,
  \dodoi{10.1017/S1743921309032025}

\bibitem[{{Irwin} {et~al.}(2008){Irwin}, {Hodgkin}, {Aigrain}, {Bouvier},
  {Hebb}, {Irwin}, \& {Moraux}}]{irwin.etal:2008}
{Irwin}, J., {Hodgkin}, S., {Aigrain}, S., {et~al.} 2008, \mnras, 384, 675,
  \dodoi{10.1111/j.1365-2966.2007.12725.x}

\bibitem[{{James} {et~al.}(2000){James}, {Jardine}, {Jeffries}, {Randich},
  {Collier Cameron}, \& {Ferreira}}]{james.etal:2000}
{James}, D.~J., {Jardine}, M.~M., {Jeffries}, R.~D., {et~al.} 2000, \mnras,
  318, 1217, \dodoi{10.1046/j.1365-8711.2000.03838.x}

\bibitem[{{Jeffries} {et~al.}(2011){Jeffries}, {Jackson}, {Briggs}, {Evans}, \&
  {Pye}}]{jeffries.etal:2011}
{Jeffries}, R.~D., {Jackson}, R.~J., {Briggs}, K.~R., {Evans}, P.~A., \& {Pye},
  J.~P. 2011, \mnras, 411, 2099, \dodoi{10.1111/j.1365-2966.2010.17848.x}

\bibitem[{{Jeffries} {et~al.}(2017){Jeffries}, {Jackson}, {Franciosini}, {Rand
  ich}, {Barrado}, {Frasca}, {Klutsch}, {Lanzafame}, {Prisinzano}, {Sacco},
  {Gilmore}, {Vallenari}, {Alfaro}, {Koposov}, {Pancino}, {Bayo}, {Casey},
  {Costado}, {Damiani}, {Hourihane}, {Lewis}, {Jofre}, {Magrini}, {Monaco},
  {Morbidelli}, {Worley}, {Zaggia}, \& {Zwitter}}]{jeffries.etal:2017}
{Jeffries}, R.~D., {Jackson}, R.~J., {Franciosini}, E., {et~al.} 2017, \mnras,
  464, 1456, \dodoi{10.1093/mnras/stw2458}

\bibitem[{{Kawaler}(1988)}]{kawaler:1988}
{Kawaler}, S.~D. 1988, \apj, 333, 236, \dodoi{10.1086/166740}

\bibitem[{{Koenigl}(1991)}]{koenigl:1991}
{Koenigl}, A. 1991, \apjl, 370, L39, \dodoi{10.1086/185972}

\bibitem[{{Kraft}(1967)}]{kraft:1967}
{Kraft}, R.~P. 1967, \apj, 150, 551, \dodoi{10.1086/149359}

\bibitem[{{Krishnamurthi} {et~al.}(1997){Krishnamurthi}, {Pinsonneault},
  {Barnes}, \& {Sofia}}]{krishnamurthi.etal:1997}
{Krishnamurthi}, A., {Pinsonneault}, M.~H., {Barnes}, S., \& {Sofia}, S. 1997,
  \apj, 480, 303, \dodoi{10.1086/303958}

\bibitem[{{Lanzafame} \& {Spada}(2015)}]{lanzafame.spada:2015}
{Lanzafame}, A.~C., \& {Spada}, F. 2015, \aap, 584, A30,
  \dodoi{10.1051/0004-6361/201526770}

\bibitem[{{Lehmann} {et~al.}(2019){Lehmann}, {Hussain}, {Jardine}, {Mackay}, \&
  {Vidotto}}]{lehman.etal:2019}
{Lehmann}, L.~T., {Hussain}, G.~A.~J., {Jardine}, M.~M., {Mackay}, D.~H., \&
  {Vidotto}, A.~A. 2019, \mnras, 483, 5246, \dodoi{10.1093/mnras/sty3362}

\bibitem[{{Maeder}(2009)}]{maeder:2009}
{Maeder}, A. 2009, {Physics, Formation and Evolution of Rotating Stars},
  \dodoi{10.1007/978-3-540-76949-1}

\bibitem[{{Maeder} \& {Meynet}(2000)}]{maeder.meynet:2000a}
{Maeder}, A., \& {Meynet}, G. 2000, \aap, 361, 159

\bibitem[{{Maeder} \& {Zahn}(1998)}]{maeder.zahn:1998}
{Maeder}, A., \& {Zahn}, J.-P. 1998, \aap, 334, 1000

\bibitem[{{Mamajek} \& {Hillenbrand}(2008)}]{mamajek.hillenbrand:2008}
{Mamajek}, E.~E., \& {Hillenbrand}, L.~A. 2008, \apj, 687, 1264,
  \dodoi{10.1086/591785}

\bibitem[{{Marsden} {et~al.}(2011){Marsden}, {Jardine}, {Ram{\'\i}rez
  V{\'e}lez}, {Alecian}, {Brown}, {Carter}, {Donati}, {Dunstone}, {Hart},
  {Semel}, \& {Waite}}]{marsden.etal:2011}
{Marsden}, S.~C., {Jardine}, M.~M., {Ram{\'\i}rez V{\'e}lez}, J.~C., {et~al.}
  2011, \mnras, 413, 1922, \dodoi{10.1111/j.1365-2966.2011.18367.x}

\bibitem[{{Mathis} \& {Zahn}(2004)}]{mathis.zahn:2004}
{Mathis}, S., \& {Zahn}, J.~P. 2004, \aap, 425, 229,
  \dodoi{10.1051/0004-6361:20040278}

\bibitem[{{Matt} \& {Pudritz}(2005)}]{matt.pudritz:2005}
{Matt}, S., \& {Pudritz}, R.~E. 2005, \apjl, 632, L135, \dodoi{10.1086/498066}

\bibitem[{{Matt} \& {Pudritz}(2008{\natexlab{a}})}]{matt.pudritz:2008a}
---. 2008{\natexlab{a}}, \apj, 678, 1109, \dodoi{10.1086/533428}

\bibitem[{{Matt} \& {Pudritz}(2008{\natexlab{b}})}]{matt.pudritz:2008b}
---. 2008{\natexlab{b}}, \apj, 681, 391, \dodoi{10.1086/587453}

\bibitem[{{Matt} {et~al.}(2015){Matt}, {Brun}, {Baraffe}, {Bouvier}, \&
  {Chabrier}}]{matt.etal:2015}
{Matt}, S.~P., {Brun}, A.~S., {Baraffe}, I., {Bouvier}, J., \& {Chabrier}, G.
  2015, \apj, 799, L23, \dodoi{10.1088/2041-8205/799/2/L23}

\bibitem[{{Matt} {et~al.}(2012{\natexlab{a}}){Matt}, {MacGregor},
  {Pinsonneault}, \& {Greene}}]{matt.etal:2012}
{Matt}, S.~P., {MacGregor}, K.~B., {Pinsonneault}, M.~H., \& {Greene}, T.~P.
  2012{\natexlab{a}}, \apjl, 754, L26, \dodoi{10.1088/2041-8205/754/2/L26}

\bibitem[{{Matt} {et~al.}(2012{\natexlab{b}}){Matt}, {Pinz{\'o}n}, {Greene}, \&
  {Pudritz}}]{matt.etal:2012b}
{Matt}, S.~P., {Pinz{\'o}n}, G., {Greene}, T.~P., \& {Pudritz}, R.~E.
  2012{\natexlab{b}}, \apj, 745, 101, \dodoi{10.1088/0004-637X/745/1/101}

\bibitem[{{Meibom} {et~al.}(2015){Meibom}, {Barnes}, {Platais}, {Gilliland},
  {Latham}, \& {Mathieu}}]{meibom.etal:2015}
{Meibom}, S., {Barnes}, S.~A., {Platais}, I., {et~al.} 2015, \nat, 517, 589,
  \dodoi{10.1038/nature14118}

\bibitem[{{Meibom} {et~al.}(2011){Meibom}, {Mathieu}, {Stassun}, {Liebesny}, \&
  {Saar}}]{meibom.etal:2011}
{Meibom}, S., {Mathieu}, R.~D., {Stassun}, K.~G., {Liebesny}, P., \& {Saar},
  S.~H. 2011, \apj, 733, 115, \dodoi{10.1088/0004-637X/733/2/115}

\bibitem[{{Meynet} \& {Maeder}(2000)}]{meynet.maeder:2000}
{Meynet}, G., \& {Maeder}, A. 2000, \aap, 361, 101

\bibitem[{{Moraux} {et~al.}(2013){Moraux}, {Artemenko}, {Bouvier}, {Irwin},
  {Ibrahimov}, {Magakian}, {Grankin}, {Nikogossian}, {Cardoso}, {Hodgkin},
  {Aigrain}, \& {Movsessian}}]{moraux.etal:2013}
{Moraux}, E., {Artemenko}, S., {Bouvier}, J., {et~al.} 2013, \aap, 560, A13,
  \dodoi{10.1051/0004-6361/201321508}

\bibitem[{{Mullan} \& {MacDonald}(2001)}]{mullan.macdonald:2001}
{Mullan}, D.~J., \& {MacDonald}, J. 2001, \apj, 559, 353,
  \dodoi{10.1086/322336}

\bibitem[{{N{\'u}{\~n}ez} {et~al.}(2015){N{\'u}{\~n}ez}, {Ag{\"u}eros},
  {Covey}, {Hartman}, {Kraus}, {Bowsher}, {Douglas}, {L{\'o}pez-Morales},
  {Pooley}, {Posselt}, {Saar}, \& {West}}]{nunez.etal:2015}
{N{\'u}{\~n}ez}, A., {Ag{\"u}eros}, M.~A., {Covey}, K.~R., {et~al.} 2015, \apj,
  809, 161, \dodoi{10.1088/0004-637X/809/2/161}

\bibitem[{{Palacios} {et~al.}(2003){Palacios}, {Talon}, {Charbonnel}, \&
  {Forestini}}]{palacios.etal:2003}
{Palacios}, A., {Talon}, S., {Charbonnel}, C., \& {Forestini}, M. 2003, \aap,
  399, 603, \dodoi{10.1051/0004-6361:20021759}

\bibitem[{{Paxton} {et~al.}(2011){Paxton}, {Bildsten}, {Dotter}, {Herwig},
  {Lesaffre}, \& {Timmes}}]{paxton.etal:2011}
{Paxton}, B., {Bildsten}, L., {Dotter}, A., {et~al.} 2011, \apjs, 192, 3,
  \dodoi{10.1088/0067-0049/192/1/3}

\bibitem[{{Paxton} {et~al.}(2013){Paxton}, {Cantiello}, {Arras}, {Bildsten},
  {Brown}, {Dotter}, {Mankovich}, {Montgomery}, {Stello}, {Timmes}, \&
  {Townsend}}]{paxton.etal:2013}
{Paxton}, B., {Cantiello}, M., {Arras}, P., {et~al.} 2013, \apjs, 208, 4,
  \dodoi{10.1088/0067-0049/208/1/4}

\bibitem[{{Paxton} {et~al.}(2015){Paxton}, {Marchant}, {Schwab}, {Bauer},
  {Bildsten}, {Cantiello}, {Dessart}, {Farmer}, {Hu}, {Langer}, {Townsend},
  {Townsley}, \& {Timmes}}]{paxton.etal:2015}
{Paxton}, B., {Marchant}, P., {Schwab}, J., {et~al.} 2015, \apjs, 220, 15,
  \dodoi{10.1088/0067-0049/220/1/15}

\bibitem[{{Paxton} {et~al.}(2018){Paxton}, {Schwab}, {Bauer}, {Bildsten},
  {Blinnikov}, {Duffell}, {Farmer}, {Goldberg}, {Marchant}, {Sorokina},
  {Thoul}, {Townsend}, \& {Timmes}}]{paxton.etal:2018}
{Paxton}, B., {Schwab}, J., {Bauer}, E.~B., {et~al.} 2018, \apjs, 234, 34,
  \dodoi{10.3847/1538-4365/aaa5a8}

\bibitem[{{Paxton} {et~al.}(2019){Paxton}, {Smolec}, {Gautschy}, {Bildsten},
  {Cantiello}, {Dotter}, {Farmer}, {Goldberg}, {Jermyn}, {Kanbur}, {Marchant},
  {Schwab}, {Thoul}, {Townsend}, {Wolf}, {Zhang}, \&
  {Timmes}}]{paxton.etal:2019}
{Paxton}, B., {Smolec}, R., {Gautschy}, A., {et~al.} 2019, arXiv e-prints.
\newblock \doarXiv{1903.01426}

\bibitem[{{Pietrinferni} {et~al.}(2004){Pietrinferni}, {Cassisi}, {Salaris}, \&
  {Castelli}}]{pietrinferni.etal:2004}
{Pietrinferni}, A., {Cassisi}, S., {Salaris}, M., \& {Castelli}, F. 2004, \apj,
  612, 168, \dodoi{10.1086/422498}

\bibitem[{{Pinsonneault} {et~al.}(1989){Pinsonneault}, {Kawaler}, {Sofia}, \&
  {Demarque}}]{pinsonneault.etal:1989}
{Pinsonneault}, M.~H., {Kawaler}, S.~D., {Sofia}, S., \& {Demarque}, P. 1989,
  \apj, 338, 424, \dodoi{10.1086/167210}

\bibitem[{{Rebull} {et~al.}(2018){Rebull}, {Stauffer}, {Cody}, {Hillenbrand},
  {David}, \& {Pinsonneault}}]{rebull.etal:2018}
{Rebull}, L.~M., {Stauffer}, J.~R., {Cody}, A.~M., {et~al.} 2018, \aj, 155,
  196, \dodoi{10.3847/1538-3881/aab605}

\bibitem[{{Rebull} {et~al.}(2016){Rebull}, {Stauffer}, {Bouvier}, {Cody},
  {Hillenbrand}, {Soderblom}, {Valenti}, {Barrado}, {Bouy}, {Ciardi},
  {Pinsonneault}, {Stassun}, {Micela}, {Aigrain}, {Vrba}, {Somers},
  {Christiansen}, {Gillen}, \& {Collier Cameron}}]{rebull.etal:2016}
{Rebull}, L.~M., {Stauffer}, J.~R., {Bouvier}, J., {et~al.} 2016, \aj, 152,
  113, \dodoi{10.3847/0004-6256/152/5/113}

\bibitem[{{Reiners} \& {Basri}(2007)}]{reiners.basri:2007}
{Reiners}, A., \& {Basri}, G. 2007, \apj, 656, 1121, \dodoi{10.1086/510304}

\bibitem[{{R{\'e}ville} {et~al.}(2015){R{\'e}ville}, {Brun}, {Matt},
  {Strugarek}, \& {Pinto}}]{reville.etal:2015}
{R{\'e}ville}, V., {Brun}, A.~S., {Matt}, S.~P., {Strugarek}, A., \& {Pinto},
  R.~F. 2015, \apj, 798, 116, \dodoi{10.1088/0004-637X/798/2/116}

\bibitem[{{Rogers}(2015)}]{rogers:2015}
{Rogers}, T.~M. 2015, \apjl, 815, L30, \dodoi{10.1088/2041-8205/815/2/L30}

\bibitem[{{Rogers} \& {McElwaine}(2017)}]{rogers.mcelwaine:2017}
{Rogers}, T.~M., \& {McElwaine}, J.~N. 2017, \apjl, 848, L1,
  \dodoi{10.3847/2041-8213/aa8d13}

\bibitem[{{Schatzman}(1962)}]{schatzman:1962}
{Schatzman}, E. 1962, Annales d'Astrophysique, 25, 18

\bibitem[{{See} {et~al.}(2020){See}, {Lehmann}, {Matt}, \&
  {Finley}}]{see.etal:2020}
{See}, V., {Lehmann}, L., {Matt}, S.~P., \& {Finley}, A.~J. 2020, \apj, 894,
  69, \dodoi{10.3847/1538-4357/ab7918}

\bibitem[{{See} {et~al.}(2016){See}, {Jardine}, {Vidotto}, {Donati}, {Boro
  Saikia}, {Bouvier}, {Fares}, {Folsom}, {Gregory}, {Hussain}, {Jeffers},
  {Marsden}, {Morin}, {Moutou}, {do Nascimento}, {Petit}, \&
  {Waite}}]{see.etal:2016}
{See}, V., {Jardine}, M., {Vidotto}, A.~A., {et~al.} 2016, \mnras, 462, 4442,
  \dodoi{10.1093/mnras/stw2010}

\bibitem[{{See} {et~al.}(2019){See}, {Matt}, {Finley}, {Folsom}, {Boro Saikia},
  {Donati}, {Fares}, {H{\'e}brard}, {Jardine}, {Jeffers}, {Marsden}, {Mengel},
  {Morin}, {Petit}, {Vidotto}, {Waite}, \& {BCool
  Collaboration}}]{see.etal:2019}
{See}, V., {Matt}, S.~P., {Finley}, A.~J., {et~al.} 2019, \apj, 886, 120,
  \dodoi{10.3847/1538-4357/ab46b2}

\bibitem[{{Sestito} \& {Randich}(2005)}]{sestito.randich:2005}
{Sestito}, P., \& {Randich}, S. 2005, \aap, 442, 615,
  \dodoi{10.1051/0004-6361:20053482}

\bibitem[{{Shu} {et~al.}(1994){Shu}, {Najita}, {Ostriker}, {Wilkin}, {Ruden},
  \& {Lizano}}]{shu.etal:1994}
{Shu}, F., {Najita}, J., {Ostriker}, E., {et~al.} 1994, \apj, 429, 781,
  \dodoi{10.1086/174363}

\bibitem[{{Skumanich}(1972)}]{skumanich:1972}
{Skumanich}, A. 1972, \apj, 171, 565, \dodoi{10.1086/151310}

\bibitem[{{Soderblom} {et~al.}(1993){Soderblom}, {Jones}, {Balachandran},
  {Stauffer}, {Duncan}, {Fedele}, \& {Hudon}}]{soderblom.etal:1993}
{Soderblom}, D.~R., {Jones}, B.~F., {Balachandran}, S., {et~al.} 1993, \aj,
  106, 1059, \dodoi{10.1086/116704}

\bibitem[{{Somers} {et~al.}(2020){Somers}, {Cao}, \&
  {Pinsonneault}}]{somers.cao.pinsonneault:2020}
{Somers}, G., {Cao}, L., \& {Pinsonneault}, M.~H. 2020, \apj, 891, 29,
  \dodoi{10.3847/1538-4357/ab722e}

\bibitem[{{Somers} \& {Pinsonneault}(2014)}]{somers.pinsonneault:2014}
{Somers}, G., \& {Pinsonneault}, M.~H. 2014, \apj, 790, 72,
  \dodoi{10.1088/0004-637X/790/1/72}

\bibitem[{{Somers} \&
  {Pinsonneault}(2015{\natexlab{a}})}]{somers.pinsonneault:2015b}
---. 2015{\natexlab{a}}, \apj, 807, 174, \dodoi{10.1088/0004-637X/807/2/174}

\bibitem[{{Somers} \&
  {Pinsonneault}(2015{\natexlab{b}})}]{somers.pinsonneault:2015}
---. 2015{\natexlab{b}}, \mnras, 449, 4131, \dodoi{10.1093/mnras/stv630}

\bibitem[{{Somers} \& {Pinsonneault}(2016)}]{somers.pinsonneault:2016}
---. 2016, \apj, 829, 32, \dodoi{10.3847/0004-637X/829/1/32}

\bibitem[{{Somers} {et~al.}(2017){Somers}, {Stauffer}, {Rebull}, {Cody}, \&
  {Pinsonneault}}]{somers.etal:2017}
{Somers}, G., {Stauffer}, J., {Rebull}, L., {Cody}, A.~M., \& {Pinsonneault},
  M. 2017, \apj, 850, 134, \dodoi{10.3847/1538-4357/aa93ed}

\bibitem[{{Spada} \& {Lanzafame}(2020)}]{spada.lanzafame:2020}
{Spada}, F., \& {Lanzafame}, A.~C. 2020, \aap, 636, A76,
  \dodoi{10.1051/0004-6361/201936384}

\bibitem[{{Th{\'e}venin} {et~al.}(2017){Th{\'e}venin}, {Oreshina}, {Baturin},
  {Gorshkov}, {Morel}, \& {Provost}}]{thevenin.etal:2017}
{Th{\'e}venin}, F., {Oreshina}, A.~V., {Baturin}, V.~A., {et~al.} 2017, \aap,
  598, A64, \dodoi{10.1051/0004-6361/201629385}

\bibitem[{{van Saders} {et~al.}(2016){van Saders}, {Ceillier}, {Metcalfe},
  {Silva Aguirre}, {Pinsonneault}, {Garc{\'\i}a}, {Mathur}, \&
  {Davies}}]{vansaders.etal:2016}
{van Saders}, J.~L., {Ceillier}, T., {Metcalfe}, T.~S., {et~al.} 2016, \nat,
  529, 181, \dodoi{10.1038/nature16168}

\bibitem[{{van Saders} \& {Pinsonneault}(2013)}]{vansaders.pinsonneault:2013}
{van Saders}, J.~L., \& {Pinsonneault}, M.~H. 2013, \apj, 776, 67,
  \dodoi{10.1088/0004-637X/776/2/67}

\bibitem[{{van Saders} {et~al.}(2019){van Saders}, {Pinsonneault}, \&
  {Barbieri}}]{vansaders.pinsonneault.barbieri:2019}
{van Saders}, J.~L., {Pinsonneault}, M.~H., \& {Barbieri}, M. 2019, \apj, 872,
  128, \dodoi{10.3847/1538-4357/aafafe}

\bibitem[{{Venn} {et~al.}(2002){Venn}, {Brooks}, {Lambert}, {Lemke}, {Langer},
  {Lennon}, \& {Keenan}}]{venn.etal:2002}
{Venn}, K.~A., {Brooks}, A.~M., {Lambert}, D.~L., {et~al.} 2002, \apj, 565,
  571, \dodoi{10.1086/324435}

\bibitem[{{Vidotto} {et~al.}(2014){Vidotto}, {Gregory}, {Jardine}, {Donati},
  {Petit}, {Morin}, {Folsom}, {Bouvier}, {Cameron}, {Hussain}, {Marsden},
  {Waite}, {Fares}, {Jeffers}, \& {do Nascimento}}]{vidotto.etal:2014}
{Vidotto}, A.~A., {Gregory}, S.~G., {Jardine}, M., {et~al.} 2014, \mnras, 441,
  2361, \dodoi{10.1093/mnras/stu728}

\bibitem[{{von Zeipel}(1924)}]{vonzeipel:1924}
{von Zeipel}, H. 1924, \mnras, 84, 665, \dodoi{10.1093/mnras/84.9.665}

\bibitem[{{Waite} {et~al.}(2015){Waite}, {Marsden}, {Carter}, {Petit},
  {Donati}, {Jeffers}, \& {Boro Saikia}}]{waite.etal:2015}
{Waite}, I.~A., {Marsden}, S.~C., {Carter}, B.~D., {et~al.} 2015, \mnras, 449,
  8, \dodoi{10.1093/mnras/stv006}

\bibitem[{{Weber} \& {Davis}(1967)}]{weber.davis:1967}
{Weber}, E.~J., \& {Davis}, Leverett, J. 1967, \apj, 148, 217,
  \dodoi{10.1086/149138}

\bibitem[{{Wright} \& {Drake}(2016)}]{wright.drake:2016}
{Wright}, N.~J., \& {Drake}, J.~J. 2016, \nat, 535, 526,
  \dodoi{10.1038/nature18638}

\bibitem[{{Wright} {et~al.}(2011){Wright}, {Drake}, {Mamajek}, \&
  {Henry}}]{wright.etal:2011}
{Wright}, N.~J., {Drake}, J.~J., {Mamajek}, E.~E., \& {Henry}, G.~W. 2011,
  \apj, 743, 48, \dodoi{10.1088/0004-637X/743/1/48}

\bibitem[{{Wright} {et~al.}(2018){Wright}, {Newton}, {Williams}, {Drake}, \&
  {Yadav}}]{wright.etal:2018}
{Wright}, N.~J., {Newton}, E.~R., {Williams}, P. K.~G., {Drake}, J.~J., \&
  {Yadav}, R.~K. 2018, \mnras, 479, 2351, \dodoi{10.1093/mnras/sty1670}

\bibitem[{{Zahn}(1992)}]{zahn:1992}
{Zahn}, J.~P. 1992, \aap, 265, 115

\end{thebibliography}

\bibliographystyle{aasjournal}



\end{document}